\documentclass[12pt,preprint]{aastex}

\input psfig.sty

\begin{document}

\title{Obscured star formation in the central region of 
the dwarf galaxy NGC~5253\footnote{Based on observations with the NASA/ESA 
Hubble
Space Telescope, obtained at the Space Telescope Science Institute, which
is operated by the Association of Universities for Research in Astronomy,
Inc. under NASA contract No. NAS5-26555.}}

\author{Almudena Alonso-Herrero\altaffilmark{2}}

\affil{Steward Observatory, The University of Arizona, 933 N. Cherry Ave., 
Tucson, AZ  85721} 

\and
\affil{Departamento de Astrof\'{\i}sica Molecular
e Infrarroja, IEM, Consejo Superior de Investigaciones Cient\'{\i}ficas,
Serrano 113b, 28006 Madrid, Spain}

\altaffiltext{2}{Visiting Astronomer at the Infrared
Telescope Facility, which is operated by the
University of Hawaii under
    Cooperative Agreement no. NCC 5-538 with the National
    Aeronautics and Space Administration, Office of Space
    Science, Planetary Astronomy Program.}

\author{Toshinobu Takagi}
\affil{Centre for Astrophysics and Planetary Science,
  University of Kent, Canterbury, Kent, CT2 7NR, UK}
\author{Andrew J. Baker}
\affil{Max--Planck--Institut f{\" u}r extraterrestrische Physik,
Postfach 1312, D--85741 Garching, Germany}
\author{George H. Rieke, Marcia J. Rieke}
\affil{Steward Observatory, University of Arizona, 933 N. 
Cherry Ave, Tucson, AZ85721}
\author{Masatoshi Imanishi\altaffilmark{2}}
\affil{National Astronomical Observatory
 2-21-1 Osawa, Mitaka, Tokyo 181-8588, Japan}
\and
\author{Nick Z. Scoville}
\affil{California Institute of Technology, Pasadena, CA 91125}

\begin{abstract}

We present {\it HST}/NICMOS observations ($1.1-2.2\,\mu$m)  and 
$1.9-4.1\,\mu$m spectroscopy of the central region of the dwarf
galaxy NGC~5253. The 
{\it HST}/NICMOS observations reveal the presence of a nuclear 
double star cluster 
separated by $\simeq 0.3-0.4\arcsec$ or $6-8$\,pc (for a distance
$d=4.1\,$Mpc). 
The double star cluster, also a bright double source of 
Pa$\alpha$ emission, appears to be coincident 
with the double radio nebula detected at 1.3\,cm. The eastern near-infrared 
star cluster (C1) is identified with the youngest optical cluster,
whereas the western star cluster (C2), although it 
is almost completely obscured in the optical, becomes
the brightest star cluster in the central 
region of NGC~5253 at wavelengths longer than $2\,\mu$m.
Both clusters are extremely 
young with ages of approximately 3.5 million years old. C2 is
more massive than C1 by a factor of 6 to 20 
($M_{\rm C2} = 7.7 \times 10^5 - 2.6 \times 10^6\,{\rm M}_\odot$,
for a Salpeter IMF in the mass range 
$0.1-100\,{\rm M}_\odot$).  Analysis of the circumnuclear
spectrum excluding C1 and C2, as well as of a number of other
near-infrared selected 
clusters with a range of (young) ages, suggests that the star 
formation was triggered across the central regions of the galaxy.  
We have also modelled 
the nuclear UV to mid-infrared 
spectral energy distribution (SED) of NGC~5253 and found that the 
infrared part is well modelled with a highly obscured 
($A_V= 17\,$mag) young starburst with a stellar 
mass consistent with our photometric estimates for C1 and C2. 
The SED model predicts a moderately bright 
polycyclic aromatic hydrocarbon  (PAH) feature at $3.3\,\mu$m  that 
is not detected in our nuclear 
$L$-band spectrum. NGC\,5253's low metallicity
and a top-heavy IMF likely combine to suppress the $3.3\,{\rm \mu m}$ PAH
emission that is commonly seen in more massive starburst systems.

\end{abstract}


\section{Introduction}

It has long been recognized that the starburst occurring in the central
region of NGC~5253 must be among  the youngest in the local 
Universe, and that this star forming process 
may be the first to have occurred in this galaxy in the last $\simeq 10^8\,$yr
(van den Bergh 1980; Moorwood \& Glass 1982; Rieke, Lebofsky, 
\& Walker 1988). 
NGC~5253 appears to have been a metal poor dwarf elliptical 
before the current burst of star formation, and after
the current burst of star formation it may evolve
back to a dwarf elliptical galaxy with a bright core such as those
in the Virgo and Fornax clusters (Caldwell \& Phillips 
1989). 

One of the longstanding questions is the triggering mechanism 
for the central starburst of NGC~5253.  It has been suggested that the 
nuclear burst of star formation was triggered as the result of an 
interaction with the neighboring galaxy M83 (van den Bergh 1980; 
Caldwell \& Phillips 1989). Recently Meier, Turner, \& Beck 
(2002) have found that only the giant molecular cloud 
in the dust lane
intersecting the galaxy, among 
the five detected in CO(2-1),  is sufficiently nearby to be
associated with the central starburst. More interestingly, 
these authors have shown that 
gas is falling into the nucleus of this galaxy, suggesting
that this may be the mechanism for fuelling the starburst. 

Evidence for an extremely young and powerful 
burst of star formation 
in the central region of NGC~5253 comes from different observational 
clues. The nuclear radio and millimeter continuum 
emission are almost entirely due to thermal 
emission from  H\,{\sc ii} regions (Beck 
et al. 1996; Mohan, Anantharamaiah, \& Goss 2001; Meier et al. 2002).
Both the detection of Wolf-Rayet optical features (Walsh \& 
Roy 1987; Conti 1991) 
and the high stellar temperature ($T>35,000-40,000$\,K),  
inferred from the near-infrared 
He\,{\sc i}($1.7\,\mu$m)/Br10
line ratio (Vanzi \& Rieke 1997) and  mid-infrared emission
lines (Beck et al. 1996; Crowther et al. 1999;
Thornley et al. 2000), can only be explained with
the presence of massive stars (IMF extending to 
at least $M \simeq 40-60\,{\rm M}_\odot$, see
Rigby \& Rieke 2004).
Finally 
Cowan \& Branch (1982) searched for two historical 
supernovae in this galaxy using the VLA, 
and did not detect them, or any other radio supernovae, 
indicating that the burst must be 
very young. The star formation appears to be occurring in a relatively
small region of the galaxy on the observed H$\alpha$ emission 
(Calzetti et al. 1997) and 
the compact mid-infrared emission of this galaxy (Rieke 1976). 
Thus, NGC~5253 offers us an ideal opportunity to study the 
early stages of extragalactic 
star formation in the relatively confined nuclear region of
this galaxy.

Throughout the paper we assume a distance to NGC~5253 of
$d=4.1\,$Mpc (Sandage et al. 1994).

\section{Observations}

\subsection{{\it HST}/NICMOS Near-Infrared Imaging} 
{\it HST}/NICMOS observations of NGC~5253  were obtained in January
1998 using the NIC2 camera, with a pixel size of 0.076\arcsec\,pixel$^{-1}$. 
The filters used were F110W, F160W and F222M, roughly
equivalent to ground-based filters $J$, $H$ and $K$ respectively. In 
addition,  narrow band observations through the F187N and F190N 
filters, covering the Pa$\alpha$ ($\lambda_{\rm rest} =
1.87\,\mu$m)+ continuum 
and $1.90\,\mu$m continuum, respectively, were obtained. 

The raw images were reduced with routines from the package 
{\sc NicRed} (McLeod 1997). The main steps in the data reduction 
involve subtraction of the first readout, dark current 
subtraction on a readout-by-readout basis, correction for linearity 
and cosmic ray rejection (using FULLFIT), and flat fielding.
The flux calibration was performed using
the conversion factors derived from measurements of the standard star P330-E
during the Servicing Mission Observatory Verification (SMOV) program
(Marcia Rieke, private communication 1999).

NICMOS only provides continuum bands to the red of the emission lines. 
The extinction in the nuclear region of NGC~5253  is known
to be very high  (even at infrared wavelengths; e.g.,
Turner et al. 2003 and references therein), and a straight 
subtraction of the longer wavelength continuum image may result in an 
overcorrection of the continuum at the wavelength of the emission line. 
We fit the continuum between 
$1.60\,\mu$m and $1.90\,\mu$m 
(using the NIC2 F160W and NIC2 F190N line-free images) on a pixel-by-pixel
basis, and interpolate to  estimate the continuum at $1.87\,\mu$m. This
interpolated continuum is then subtracted from the line+continuum 
image to produce the continuum subtracted Pa$\alpha$ image.
The fully-reduced images were rotated to the usual 
orientation with north up, east to the left. They are shown in Fig.~1. 
In addition, we constructed
infrared color maps equivalent to ground-based $J-H$ (F110W - F160W) 
and $H-K$ (F160W - F222M) color maps.

\subsection{IRTF/Spex Near-Infrared $K$ and $L$-band Spectroscopy}
We have obtained simultaneous intermediate resolution $K$ ($\lambda = 1.9-
2.4\,\mu$m) and $L$ band ($\lambda = 3-4.1\,\mu$m) spectroscopy of the  
central region of NGC~5253 using Spex (Rayner et al. 2003)
at the 3\,m NASA IRTF telescope on Mauna Kea in March 2003.
We used the 0.15\arcsec\,pixel $^{-1}$ plate scale and a slit width 
of 1.6\arcsec \ which provided a spectral resolution 
of $R\simeq 600$ in the $K$-band. The slit was oriented approximately 
North-South to cover the nucleus (C1+C2, Section~3.1, Fig.~2) 
of NGC~5253 plus the bright star cluster 
C3+C3' to the southwest of the nucleus 
(see Table~1 and Fig.~2). The C4+C5 cluster (see Fig.~2) was sufficiently 
bright and pointlike at the 
guider resolution that we could use it for the tip-tilt capability, 
thus improving the spatial sampling of the data.  
The total on-source integration time was 30 minutes. The seeing conditions
as measured from standard star observations were 
approximately $0.6\arcsec$ (FWHM) in the $K$-band. 
We also obtained $KL$ spectroscopy of two comparison 
galaxies Henize 2-10 and NGC~2903 in a similar fashion.
Observations of solar-type stars were interspersed with the galaxy 
observations selected to be at similar air masses, and   
used to perform the telluric correction.

We used the data reduction package {\sc SpexTool} (Cushing,
Vacca, \& Rayner 2004)
especially developed to reduce Spex data. Briefly, we created  
calibration frames (flatfields and arcs), median combined the 
galaxy observations, flatfielded the data, extracted 
1D spectra, and performed 
the wavelength calibration using sky lines and arcs. The 1-D spectra
of the nuclear region of NGC~5253 
were extracted with apertures: 
$1.5\arcsec \times 1.6\arcsec$ and $3\arcsec \times 1.6\arcsec$. We
also extracted a $K$-band  spectrum of the C3+C3' cluster 
through a $1.5\arcsec \times 1.6\arcsec$ aperture. 

Standard stars at a similar airmass observed before and after the galaxy 
were reduced in a similar fashion and used to perform 
the correction of the galaxy spectra for atmospheric transmission. 
The flux calibration of
the galaxy spectra was obtained from the observations of the standard
stars. We compared the Spex $K$-band continuum level with the flux
measured in the NIC2 F222M images through the same 
extraction apertures and found that they agree to within 
10\%. The fully reduced Spex spectra corrected 
for the standard star blackbody temperature, flux-calibrated and 
shifted to rest frame wavelengths are shown in Fig.~3.
Line fluxes and equivalent widths (EW) 
for the main emission lines are given in Table~2.

\subsection{{\it HST}/WFPC2 UV and Optical Imaging}
Fully reduced {\it HST}/WFPC2 images of NGC~5253 through the 
F255W, F547M and F814W filters (Calzetti et al.
1997) were retrieved from the {\it HST} archive. These observations
placed the central part of the galaxy on the WP3 chip providing 
a plate scale of $\simeq 0.1\arcsec\,$ pixel$^{-1}$.  
In addition we retrieved images through the F555W and F814W filters where 
the galaxy was centered on the WP1 chip providing higher
spatial resolution. Because the 
[O\,{\sc iii}]$\lambda$5007\AA \ emission line in the central  
region of NGC~5253 is bright (e.g., Gonz\'alez-Riestra, Rego, \&
Zamorano 1987), we expect it to have a large contribution 
to the photometry through the F555W filter. We thus use 
the F555W image for comparing the optical
and near-infrared morphologies, but not for star cluster photometry.

\subsection{{\it ISO}/ISOCAM Mid-Infrared Imaging}
We obtained fully reduced {\it ISO}/ISOCAM images of NGC~5253 observed
through five 
filters: LW1 ($4.5\,\mu$m), LW2 ($6.7\,\mu$m), LW6 ($7.7\,\mu$m),
LW10 ($12\,\mu$m), and LW9 ($14.9\,\mu$m). The pixel size
is 1.5\arcsec\,pixel$^{-1}$, which provides a field of view of 
$48\arcsec \times 48\arcsec$.

\clearpage

\begin{figure*}

\epsscale{0.8}
\plotone{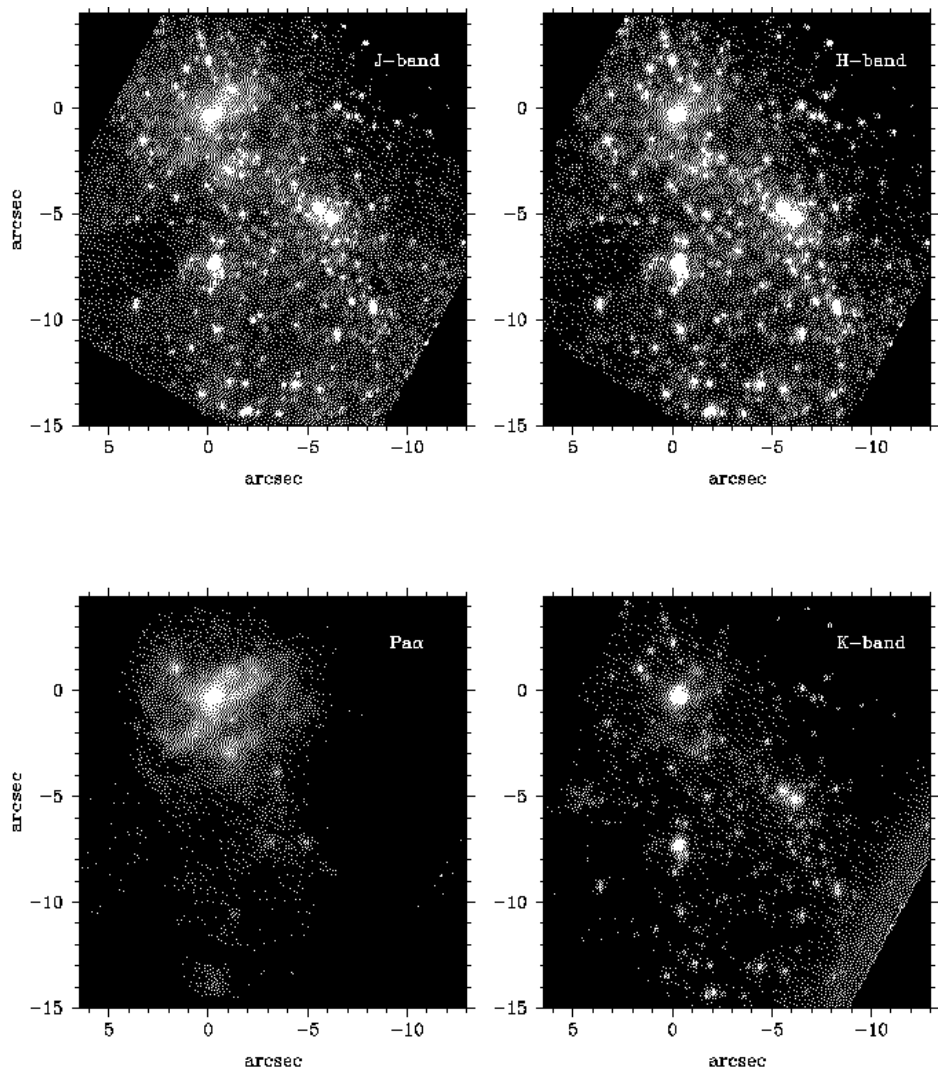}
\caption{{\it HST}/NICMOS images of the central $19\arcsec \times
19\arcsec$ of NGC~5253. Orientation is North up, East to the left. 
The continuum images are the $J$-band image (NIC2 F110W filter) on the upper 
left panel,
the $H$-band image (NIC2 F160W) on the upper right panel, 
and the $K$-band image (NIC2 F222M) on the lower right panel. 
The continuum-subtracted Pa$\alpha$ (NIC2 F187N - F190N) line image 
is on the lower left panel. The 
double star cluster C1+C2 (H1+H2) is the bright
source on the upper left of each image (see also Fig.~2).}
\end{figure*}

\begin{figure*}
\figurenum{2}
\epsscale{0.8}
\plotone{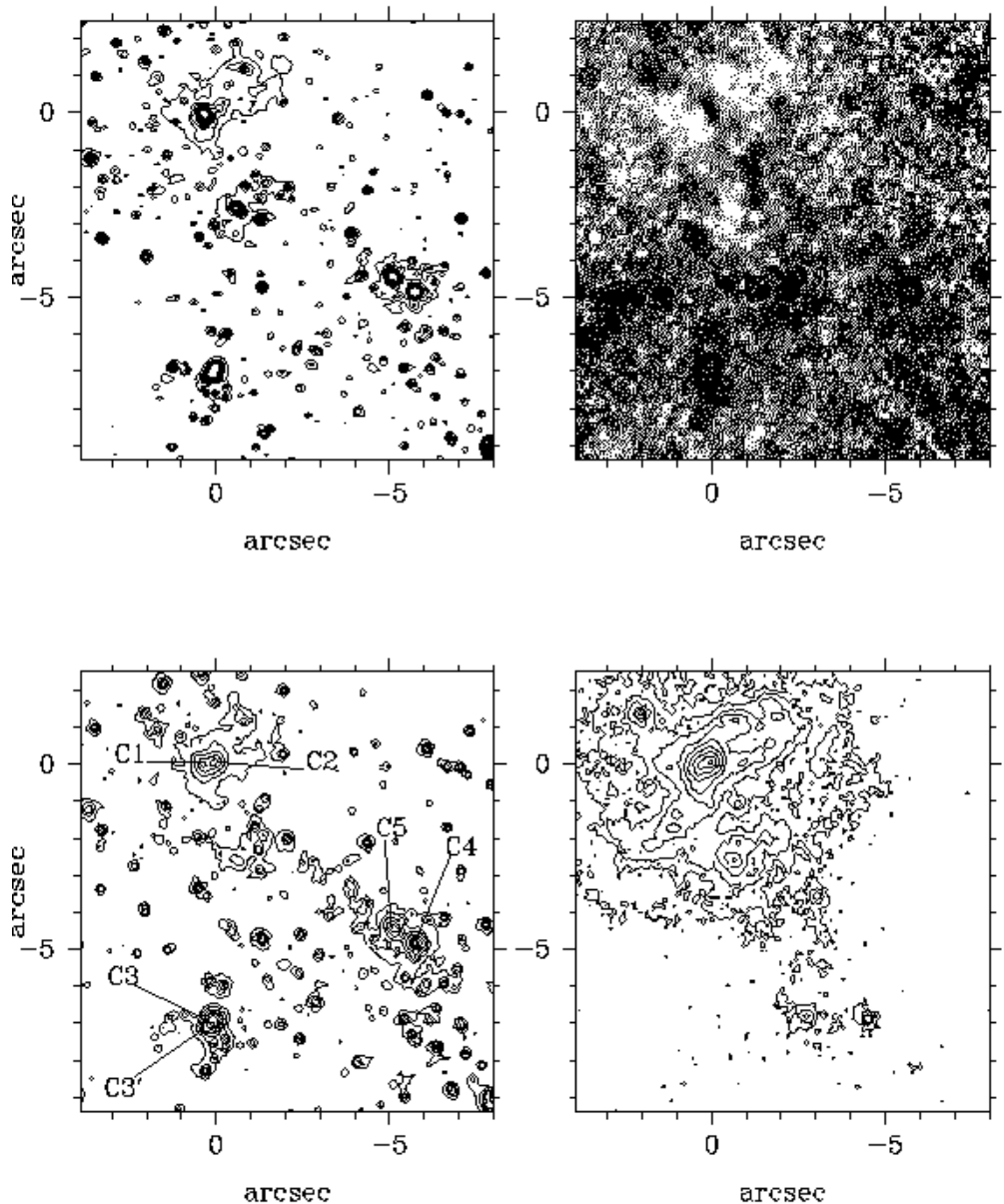}
\caption{Close-ups of the central $11.9\arcsec \times
11.9\arcsec$ of NGC~5253. Orientation is North up, East to the left. 
{\it Upper left panel:} Contours of the {\it HST}/WFPC2 F814W (WP1 chip)
image. The pixels  
have been rebinned to match the pixel size of 
the NIC2 images.
{\it Upper right panel:} 
$J-H$ color map, where dark means redder $J-H$ colors. 
{\it Lower left panel:} Contours of the 
$H$-band emission (NIC2 F160W). We mark the positions of
the five brightest $H$-band clusters.
{\it Lower right panel:} Contours of  
the continuum-subtracted Pa$\alpha$ (NIC2 F187N - F190N)
emission. }
\end{figure*}

\begin{figure*}
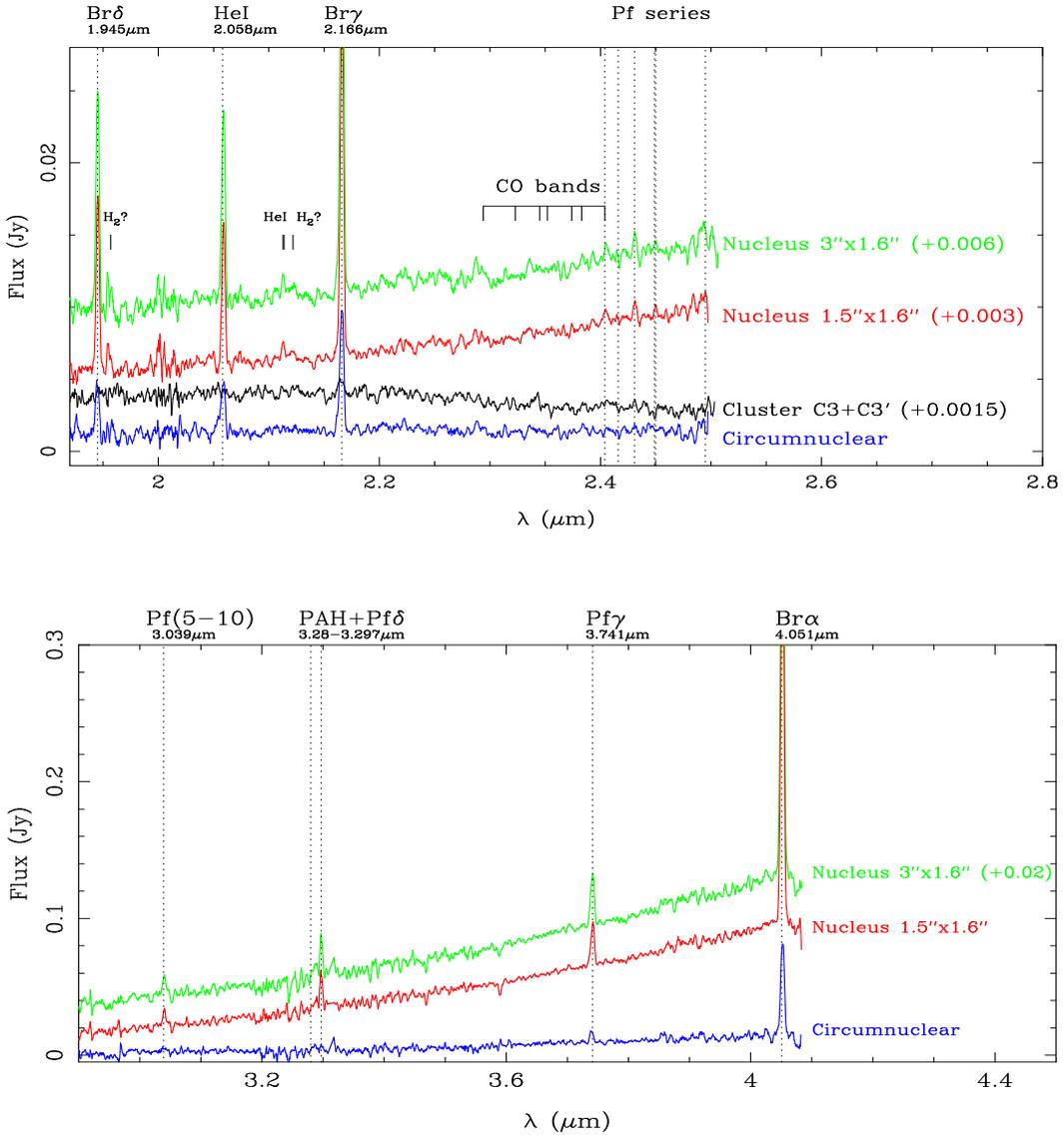

\figurenum{3}
\hspace{1cm}
\psfig{file=f3a.ps,height=7cm,width=14cm,rwidth=1cm,angle=-90}

\vspace{1cm}
\hspace{1cm}
\psfig{file=f3b.ps,height=7cm,width=14cm,angle=-90}

\caption{\footnotesize {\it Upper panel:} Spex $K$-band spectra of the nuclear
region of NGC~5253 extracted through two different apertures:
$1.5\arcsec \times 1.6\arcsec$ and $3\arcsec \times 1.6\arcsec$. In
addition we show spectra of 
a circumnuclear region (that is, excluding C1 and C2, see Section~5.1) 
and the bright cluster C3+C3' to the southwest
of the nucleus (see Fig.~2 and Table~1 for location 
with respect to the nucleus of NGC~5253). On top of the plot we show
the positions of emission lines positively identified, as well as 
tentative identifications of  H$_2$ lines
at $1.967\,\mu$m and $2.1218\,\mu$m, 
and He\,{\sc i} lines at $2.1128\,\mu$m and $2.1137\,\mu$m
(see Lumsden, Puxley,  \& Doherty 1994). We also mark 
the positions of the bandhead of the CO 
features (Kleinmann \& Hall 1986). 
For display
purposes the $K$-band spectra have been shifted along
the vertical direction by the offsets indicated in the figure.
{\it Lower panel:} 
Spex $L$-band spectra of the central
region of NGC~5253 extracted through two different apertures:
$1.5\arcsec \times 1.6\arcsec$ and $3\arcsec \times 1.6\arcsec$,
and the circumnuclear region. The spectra 
have been shifted to rest-frame wavelengths. We also mark 
the positions of bright emission lines.}
\end{figure*} 

\clearpage

\begin{deluxetable}{ccccccccl}

\tabletypesize{\scriptsize} 
\rotate
\setlength{\tabcolsep}{0.02in}
\tablecaption{Clusters in the central region of NGC~5253
with $M_H\le -11\,$mag.}
\tablehead{\colhead{Cluster}  & \colhead{$X$} &
\colhead{$Y$} & \colhead{$M_{\rm F160W}$} & \colhead{$M_{\rm F222M}$} &
\colhead{$m_{\rm F110W}-m_{\rm F160W}$} 
& \colhead{$m_{\rm F160W}-m_{\rm F222M}$} & \colhead{$\log EW({\rm Pa}
\alpha)$} & \colhead{Possible ID}\\
& \colhead{($\arcsec$)} &\colhead{($\arcsec$)} & & & & 
& \colhead{(\AA)} }

\startdata
\multicolumn{9}{c}{Central Double Cluster}\\

\hline

C2  &  0.0 &   0.0 & $-11.68\pm0.13$ & $-14.16\pm0.06$ &  $1.6\pm0.1$ &    
$2.5\pm0.1$&    $3.48\pm0.23$
& Dominant 1.3cm source\\
C1  & 0.3E &  $0.1$S & $-11.57\pm0.14$ & $-12.21\pm0.16$ &  $0.3\pm0.2$ &    
$0.6\pm0.2$&    $3.41\pm0.29$ 
& NGC5253-5, UV-12, Harris1 \\
\smallskip
& & & & & & & & Secondary 1.3cm source\\
\hline

\multicolumn{9}{c}{Other Bright Clusters}\\

\hline
C3$^*$  & 0.1E &  $7.0$S & $-12.74\pm0.16$ & $-12.38\pm0.13$ &  $0.9\pm0.2$ &  
$-0.4\pm0.2$& \nodata  & 
NGC5253-1, UV-2, Harris2\\
C3'$^*$ & 0.3E &  $7.2$S & $-11.43\pm0.11$ & $-12.72\pm0.10$ &  $1.0\pm0.1$&
$1.3\pm0.2$ & \nodata  
&\nodata \\
C3+C3' & 0.1E  & 7.1S & $-13.48\pm0.05$ & $-13.85\pm0.05$ & $1.0\pm0.1$ & 
$0.4\pm0.1$& $1.26\pm0.20$ & (0.3W,7.0S)\\

C4  &  5.7W &  $4.9$S & $-12.67\pm0.07$ & $-12.83\pm0.06$ &  $0.9\pm0.1$ &   
 $0.2\pm0.1$& \nodata   
& NGC5253-3, UV-4, Harris3 \\
C5  &  5.1W &  $4.5$S & $-11.85\pm0.15$ & $-11.53\pm0.09$ &  $0.8\pm0.2$ &   
$-0.3\pm0.2$&    $0.88\pm0.20$ 
& NGC5253-2, Harris5 \\
C6  & 0.4E &   2.5N & $-11.43\pm0.10$ & $-11.61\pm0.06$ &  $1.0\pm0.1$ &  
$0.2\pm0.1$& \nodata   
& Harris32\\
C7    & 1.6W &  13.9S & $-11.32\pm0.04$ & $-11.65\pm0.03$ &  $1.0\pm0.1$ &   
$0.3\pm0.1$ & \nodata & 
 Harris27\\
C8    & 7.9W &   9.2S & $-11.19\pm0.08$ & $-11.41\pm0.06$ &  $0.7\pm0.1$ &   
$0.2\pm0.1$ & \nodata & 
NGC5253-6?, Harris9\\
C9    & 1.3W &   4.8S & $-11.10\pm0.03$ & $-11.36\pm0.02$ &  $1.0\pm0.1$ &   
$0.3\pm0.1$ &  $1.47\pm0.09$
& \nodata\\
C10   & 4.0W &  12.8S & $-11.07\pm0.05$ & $-11.41\pm0.04$ &  $0.9\pm0.1$ &   
$0.3\pm0.1$ & \nodata 
& \nodata \\
C11   & 6.1W &  10.3S & $-11.06\pm0.06$ & $-11.39\pm0.05$ &  $1.0\pm0.1$ &   
$0.3\pm0.1$ & \nodata 
& \nodata \\
C12   & 7.8W &   9.0S & $-11.04\pm0.08$ & $-11.25\pm0.05$ &  $0.9\pm0.1$ &   
$0.2\pm0.1$ & $1.68\pm0.18$ 
& NGC5253-6?\\

\enddata
\tablecomments{$X$ and $Y$ are the measured offsets in arcseconds relative 
to the position of C2. The magnitudes, colors, and EWs are measured 
through a $0.76\arcsec$-diameter aperture (except for C3+C3', see below),
and are not corrected for extinction.
The column 'ID' indicates possible 
correspondences with sources detected at other 
wavelengths. ``NGC5253-1, 
NGC5253-2, NGC5253-3, NGC5253-5, and NGC5253-6'' are optical 
star clusters
from Calzetti et al. 1997. ``UV-2, UV-4 and UV-12'' 
are UV clusters identified by Meurer et al. 1995. Also
listed are correspondences with clusters reported
by Harris et al. (2004). The 1.3cm radio sources
are from Turner et al. 2000. The near-infrared hot spots in Davies 
et al. 1998 are named after the offsets quoted in their 
paper relative to the nucleus, where (0.3W,7.0S) corresponds to C3+C3'
and (5.8W,4.6S) to C4+C5.\\
$^*$The C3 and C3' clusters are severely blended and thus the colors 
and magnitudes
are strongly affected by deconvolution errors. We give magnitudes and 
colors for 
C3+C3' through a $1.5\arcsec$-diameter aperture.}
\end{deluxetable}

\begin{deluxetable}{cllll}
\footnotesize
\tablecaption{Line fluxes: {\it HST}/NICMOS Pa$\alpha$
measurements, and Spex $K$- and $L$-band spectroscopy 
of the nuclear region of NGC~5253.}
\tablehead{\colhead{Line}  & \colhead{$\lambda_{\rm rest}$} &
\colhead{Aperture} & \colhead{Flux} & \colhead{EW} \\
        &  \colhead{($\mu$m)} & \colhead{}& 
\colhead{(erg cm$^{-2}$ s$^{-1}$)} &\colhead{(\AA)}}
\startdata
Pa$\alpha$ & 1.877 & 
$1.5\arcsec \times 1.6\arcsec$ & $4.75\times 10^{-13}$ &  $\simeq 2000$\\
\smallskip
  &   & $3\arcsec \times 1.6\arcsec$ &
$5.96\times 10^{-13}$ & $\simeq 2200$\\
Br$\delta$ & 1.945 & 
$1.5\arcsec \times 1.6\arcsec$ & $3.40\times 10^{-14}$ & 177\\
\smallskip
  &   & $3\arcsec \times 1.6\arcsec$ &
$4.60\times 10^{-14}$ & 146\\
He{\sc i} & 2.058 & 
$1.5\arcsec \times 1.6\arcsec$ & $2.48\times 10^{-14}$ & 113\\
\smallskip
 & & $3\arcsec \times 1.6\arcsec$ & 
$3.75\times 10^{-14}$ & 116\\
Br$\gamma$ & 2.166 & $1.5\arcsec \times 1.6\arcsec$  &
$5.72 \times 10^{-14}$ & 255 \\
  &   & $3\arcsec \times 1.6\arcsec$ &
$8.25\times 10^{-14}$ & 250\\
& & circumnuclear & $2.67\times10^{-14}$ & 216\\
\smallskip
& & cluster C3+C3'  & \nodata & $<5$\\
Pf$\gamma$ & 3.741 & $1.5\arcsec \times 1.6\arcsec$ & 
$4.43\times 10^{-14}$ & 32\\

\smallskip
  &   & $3\arcsec \times 1.6\arcsec$ &
$5.90\times 10^{-14}$& 37\\
Br$\alpha$ & 4.051 &  $1.5\arcsec \times 1.6\arcsec$ & 
$3.62\times10^{-13}$ & 205\\
\smallskip
  &   & $3\arcsec \times 1.6\arcsec$ &
$4.64\times 10^{-13}$ & 235\\
\enddata
\tablecomments{Pa$\alpha$ measurements are from the narrow-band 
{\it HST}/NICMOS
imaging data. Other measurements are from the IRTF Spex spectroscopy. }
\end{deluxetable}

\clearpage

\begin{figure*}
\figurenum{4}
\epsscale{0.6}
\plotone{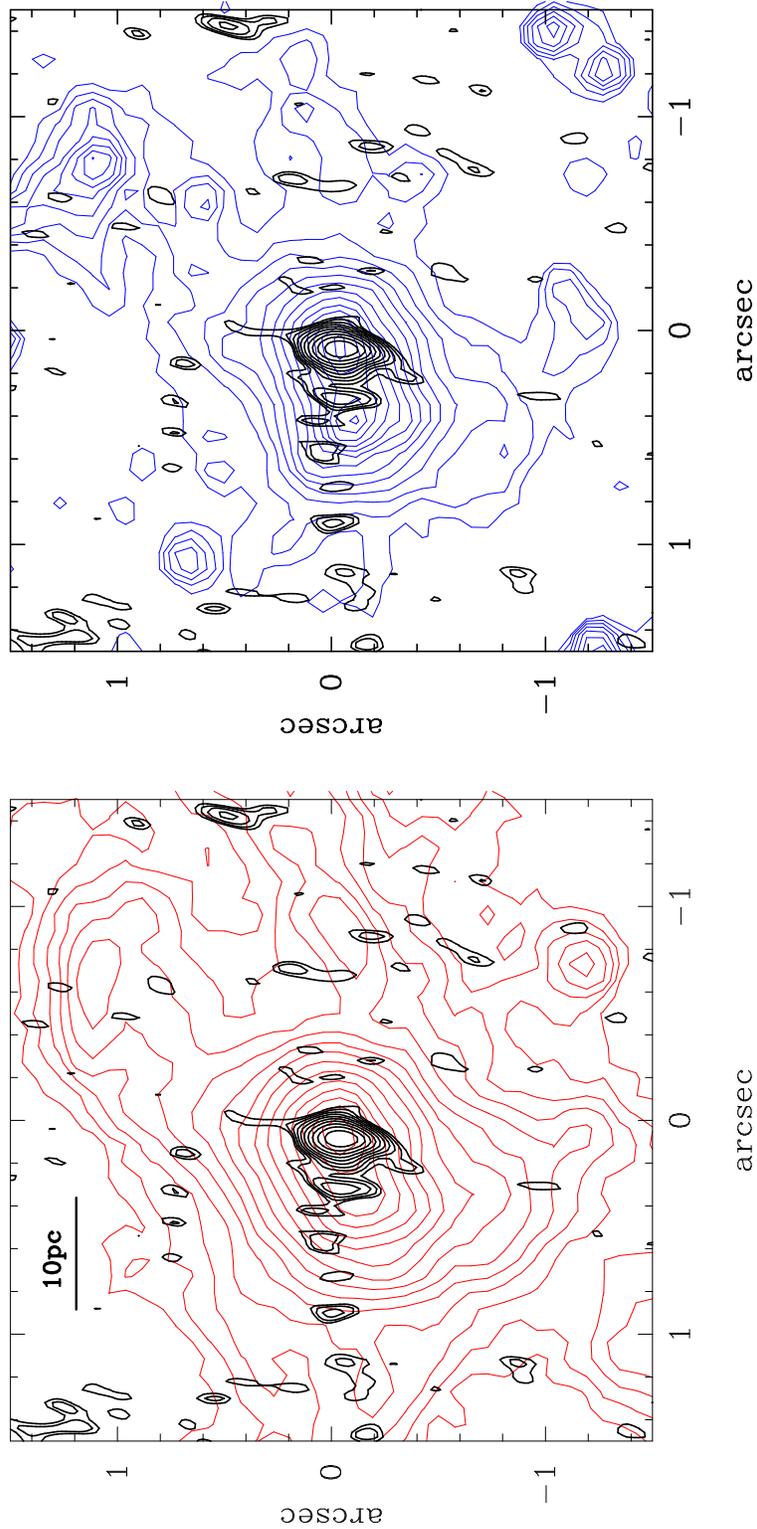}
\caption{Close-ups of the central $3\arcsec \times
3\arcsec$ of NGC~5253,  showing the two nuclear compact H\,{\sc ii} 
regions (H1+H2)/star clusters (C1+C2). The thin lines are 
the contours of the continuum-subtracted Pa$\alpha$ emission (left panel, 
in red in the electronic edition), and 
 the $1.6\,\mu$m continuum emission (right panel, in blue in the 
electronic edition).
In both panels the  thick contours are  
the 1.3cm VLA radio map from Turner et al. (2000) showing
the location of the two  radio sources. Orientation 
is North up, East to the left. }
\end{figure*} 

\begin{figure}
\figurenum{5}
\hspace{-5cm}
\psfig{file=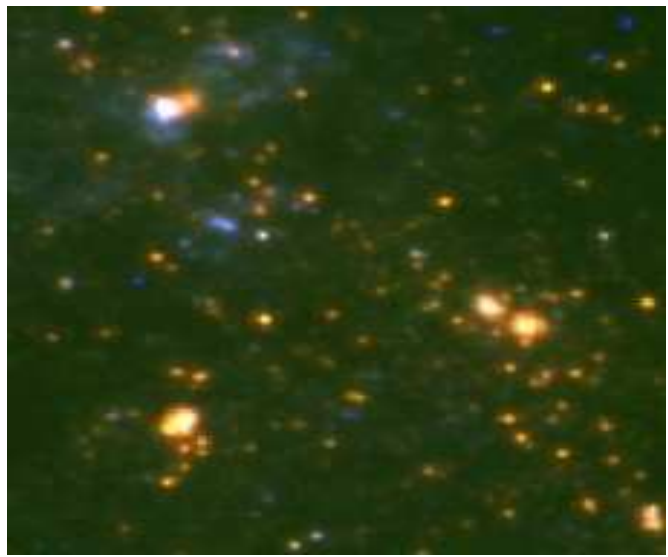,height=18cm,width=18cm,angle=0}
\vspace{-9cm}
\caption{False color image produced with the  {\it HST}/WFPC2 F555W, 
{\it HST}/NICMOS F110W, and {\it HST}/NICMOS F160W images. The 
FOV is similar to that in Fig.~2. Note the red color of the C2 cluster.}
\end{figure} 

\clearpage

\section{Photometry}

\subsection{$H$-band selected star clusters}
We choose to identify the star clusters in the central 
region of NGC~5253 using the $H$-band (NIC2 F160W 
filter) data because this wavelength represents the best compromise
between spatial resolution and reduced effects of extinction. 
We used IRAF {\sc daophot} routines and detected a total of 269 clusters. 
The positions of the clusters 
identified in the $H$-band were then used as input for {\sc daophot}
at the other four wavelengths, 1.1, 1.87, 1.9 and $2.2\,\mu$m.
Since most of the clusters appear to 
be slightly resolved, we constructed at each 
wavelength a point spread function 
(PSF) from bright sources in the images to model the 
photometry instead of using
theoretical PSFs. These PSFs are used by {\sc daophot} to model
the cluster photometry simultaneously for all the sources
detected in the field of view. The cluster photometry was obtained through a 
10\,pixel=0.76\arcsec \ ($\simeq 15\,$pc) diameter aperture.
The underlying galaxy emission was measured from annuli surrounding the 
source. In addition to the continuum photometry, 
we compute for each cluster the EW of 
Pa$\alpha$. 

Table~1 lists the photometry for the brightest star clusters 
selected in the $H$-band with absolute magnitudes
$M_{\rm F160W} \le -11$. The quoted errors are the output
of {\sc daophot}. Since most of the clusters, and in particular the 
bright ones, appear to be slightly resolved, we did not apply 
an aperture correction to the
magnitudes measured through the $0.76\arcsec$-diameter. 
The aperture corrections (for point sources) are 
$-0.18$, $-0.19$ and $-0.28$ for the F110W, F160W and F222M magnitudes, 
respectively, and 
$+0.09\,$mag for the $m_{\rm F160W}-m_{\rm F222M}$ colors, based on photometry 
of point sources generated with Tiny Tim 
(Krist et al. 1998).

\subsection{H\,{\sc ii} regions}

We identified and performed photometry of H\,{\sc ii} regions 
in the continuum subtracted Pa$\alpha$ image of 
the central $19\arcsec \times 19\arcsec$ region 
of NGC~5253.  We used the software {\sc region}, 
kindly provided by C. H. Heller (see Pleuss, Heller, \& Fricke 2000, 
and references therein for a detailed description). 

{\sc region} is a 
semiautomated method that locates and computes statistics of H\,{\sc ii} 
regions in an image, based on contouring taking into account 
the local background. We set the limit for the size of an H\,{\sc ii} 
region to 9 contiguous pixels, which at the distance of NGC~5253 
corresponds to minimum linear 
sizes (diameters) of approximately 5\,pc. Briefly, the 
H\,{\sc ii} selection
and photometry procedure is as follows.  First the program 
selects local maxima; then we set a brightness limit 
for the pixel to be included 
as part of an H\,{\sc ii} region that the pixel must 
have an intensity above the local background of at least 3 times 
the rms noise of the local background (see Rand 1992 for more 
details on the criteria employed). After identifying the H\,{\sc ii} 
region, the software measures its position, size (area), and 
Pa$\alpha$ flux by subtracting the closest local background 
(selected interactively by the user) from the observed flux. 

In Table~3 we present the photometry of H\,{\sc ii} regions with 
measured Pa$\alpha$ luminosities (not corrected for extinction) 
${\rm log}\,(L_{\rm Pa\,\alpha}/{\rm erg\,s^{-1}}) > 38$.
The diameters
listed in this table are 'equivalent diameters' computed
assuming that the area covered by the H\,{\sc ii} region is
circular.

\subsection{Nuclear Spectral 
Energy Distribution}

In deriving the nuclear spectral energy distribution (SED) 
we can only zoom in as tightly as the smallest
of our extraction apertures ($1.5\arcsec \times 1.6\arcsec$) for the 
$KL$-spectroscopy. We thus matched the photometry at other
wavelengths to this size, and paid special attention to subtracting
the underlying galaxy emission to isolate as much
as possible the nuclear emission. In what follows, we 
describe the procedure for obtaining the photometry and estimating the 
galaxy contribution. The nuclear photometry 
is summarized in Table~4.

For the photometry of the 
{\it HST}/WFPC2 and NICMOS broad-band images we simulated the 
spectroscopy aperture 
on the images, and the local background was estimated 
from annuli around the nucleus taking care that they 
did not include faint star clusters. We estimated the continuum
at $1.90\,\mu$m from the narrow-band F190N image in a similar
fashion. 

The ISOCAM images of NGC~5253 show at all five wavelengths a bright point
source superimposed on a faint extended 
component. The emission from the point source is unresolved, as  
determined from the comparison of the measured FWHM of the nuclear source 
with the sizes of theoretical 
PSFs (with sizes typically of
$2-4.5\arcsec$ FWHM depending on the
wavelength, Okumura 1998). We assume that the unresolved flux at all 
five wavelengths arises from the nuclear region. The local 
background was estimated from areas near the edges of the images. 

We also make use of mid-infrared data from the literature. 
The  mid-infrared narrow-band and broad-band fluxes of Frogel, Elias, 
\& Phillips (1982) were observed through an 8\arcsec-diameter aperture. 
This photometry is in good agreement with 
the $8-13\,\mu$m spectrophotometry  through a 
5.4\arcsec-diameter aperture of Aitken et al. (1982).
Other ground-based measurements from the literature include  
the 11.7 and $18.7\,\mu$m unresolved fluxes from Gorjian et al. (2001),
and the 10 and $21\,\mu$m fluxes through a 5.4\arcsec-diameter 
aperture of Rieke (1976) and Lebofsky \& Rieke (1979). From comparison of 
the ground-based measurements, the only discrepant point for 
the nuclear photometry is the 
8''-diameter flux at $20\,\mu$m  from Frogel et al. (1982). 
The ground-based small aperture mid-infrared fluxes 
are consistent with the ISOCAM unresolved fluxes, giving us 
confidence that most of the small aperture mid-infrared 
emission is from the 
nucleus.\footnote{We note that Rieke (1976) and
Lebofsky \& Rieke (1979) already 
remarked that NGC~5253 has a compact
$10\,\mu$m source with a much higher surface brightness 
than  other star forming galaxies in their sample. 
Gorjian et al. (2001) used Keck and  measured a size for the central
mid-infrared source of 0.62\arcsec, supporting
the fact that the mid-infrared flux originates in a very small
nuclear region.}
The IRAS fluxes (Roche et al. 1991) and the SCUBA (James et al. 2002)
data point will be taken as upper 
limits for the modeling of the nuclear SED (Section~7) as they encompass 
the entire galaxy.

\section{Morphology of the central super star cluster/H\,{\sc ii} Region}
The most striking feature revealed by 
the {\it HST}/NICMOS images is the presence of a double star cluster
(see also Turner
\& Beck 2004) in the nucleus of NGC~5253, which we term 
C1 for the eastern cluster and C2 for the western cluster (see Table~1). 
Fig.~2 shows close-ups of the central $11.9\arcsec \times
11.9\arcsec$ of NGC~5253 in the optical 
(WFPC2 F814W) and at near-infrared wavelengths. We also indicate
the locations of the five brightest clusters in the $H$-band (see 
also Table~1). 
In Fig.~4 the close-ups show the central $3\arcsec \times 3\arcsec$.

The double star cluster is most conspicuous at $1.6\,\mu$m (see 
Fig.~4), where both 
clusters show similar apparent
brightness (see Table~1). The distance between 
the two star clusters as measured at $1.6\,\mu$m is 
$0.3-0.4\arcsec$ or $6-8$\,pc for the assumed distance. 
Turner et al. (2003) have recently 
reported that the brightest source at $2.2\,\mu$m   
is offset by $\simeq 0.3\pm 0.1\arcsec$ \ to the northwest of the  
youngest optical source and peak of 
the H$\alpha$ emission ---referred to as NGC~5253-5 and UV-12 in 
Calzetti et al. (1997) and 
Meurer et al. (1995), respectively. Indeed the NICMOS F222M image
shows that the western cluster C2 dominates the emission at $2.2\,\mu$m.
We thus identify the eastern star cluster C1 
with  the peak of the H$\alpha$ emission and youngest optical star cluster
(NGC~5253-5 in Calzetti et al. 1997 notation). 
The  western star cluster C2 becomes
the brightest source at $2.2\,\mu$m (see Table 1, and Fig.~2), and 
is so reddened that it is hardly detected in the optical (see 
the WFPC2 F814W image shown in Fig.~2). This fact is also illustrated in 
the false color image of the central region of NGC~5253 shown in Fig.~5.

Bright Pa$\alpha$ emission is detected from both star clusters, and the 
two H\,{\sc ii} regions are named H2 (at the location of C2) and 
H1 (at the location of C1), see Table~3.  
The H$\alpha$ luminosity derived from the Pa$\alpha$ luminosity 
(not corrected for extinction) using the theoretical
line ratios in Hummer \& Storey 1987) of H2 indicates that 
this source has the 
typical brightness of giant H\,{\sc ii} regions, e.g., 30 
Doradus.\footnote{Note that the near-infrared 
magnitudes and EWs of C1 and C2
do not correspond to exactly the  
same spatial extent as the double H\,{\sc ii} region because
of the two (unavoidably) different methods for doing photometry of 
star clusters (approximately point sources) and H\,{\sc ii} regions
(extended sources).}

Turner, Beck, \& Ho (2000) obtained 1.3\,cm and 2\,cm VLA images of NGC~5253 
at subarcsecond resolution and detected at both wavelengths
a bright supernebula. Their 1.3\,cm map also displays a secondary peak within
the central 0.5\arcsec. Fig.~4  compares the radio and near-infrared
morphologies (Pa$\alpha$ emission and 
$1.6\,\mu$m) of the central $3\arcsec \times 3\arcsec$. Assuming that 
the brightest radio source and the C2 star cluster (also the 
H2 H\,{\sc ii} region) 
are coincident, then the secondary radio peak appears to be at the 
location of secondary H\,{\sc ii} region (H1) and at the approximate
location of the star cluster C1.  The central double cluster/H\,{\sc ii}
region appears also to be coincident with the bright mid-infrared
source detected by Gorjian, Turner, \& Beck (2001). It is likely
that the bright near-infrared cluster C2 is actually responsible 
for most of the mid-infrared emission (see Turner \& Beck 2004).

\clearpage

\begin{deluxetable}{cccccc}
\tiny
\tablecaption{H\,{\sc ii} regions in the central region of NGC~5253
with ${\rm log}\,(L_{\rm Pa\,\alpha}/{\rm erg\,s^{-1}}) > 38$.}
\tablehead{\colhead{HII}  & \colhead{$X$} &
\colhead{$Y$} & \colhead{Diameter} &
\colhead{$\log L(\rm Pa\alpha)$} & \colhead{ID}\\
\colhead{Region} & \colhead{($\arcsec$)} &\colhead{($\arcsec$)} & 
\colhead{(\arcsec)} & \colhead{(erg s$^{-1}$)}}
\startdata
\multicolumn{6}{c}{Central Double H\,{\sc ii} Region}\\
\hline
H2 & 0      & 0     & 2.0 & 38.91  & C2, Dominant 1.3cm source\\
H1 & 0.3E   & 0.1S & 1.1 & 38.42  & C1, Secondary 1.3cm source\\
   &        &       &     &        & NGC5253-5, UV-12\\
\hline
\multicolumn{6}{c}{Other H\,{\sc ii} Regions}\\
\hline
H3 & 0.8W   & 0.1S & 1.2 & 38.11 & \nodata  \\
H4 & 1.1E   & 1.7S  & 1.5 & 38.07 & \nodata \\
H5 & 0.8W   & 1.1N  & 1.1 & 38.00 & \nodata \\
H6 & 0.7W   & 2.6S  & 1.3 & 38.00  & NGC5253-4, UV-1, Harris4
\enddata
\tablecomments{$X$ and $Y$ are the measured offsets in arcseconds relative 
to the position of H2. The Pa$\alpha$ luminosities are not
corrected for extinction.
The diameters listed in this table are 'equivalent diameters' computed
assuming that the covered area by the H\,{\sc ii} region is
circular. The Pa$\alpha$ luminosities correspond to these areas, and 
are not corrected for extinction. The column 'ID' indicates possible 
correspondences with sources detected at other 
wavelengths. ``NGC5253-5/UV-12'' and ``NGC5253-4/UV-1''  
are  young star clusters 
detected in both the optical and UV, Calzetti et al. 1997 
and Meurer et al. 1995, respectively. ``Harris4'' is
an optical cluster reported by Harris et al. (2004). 
The 1.3cm radio sources
are from Turner et al. 2000.}
\end{deluxetable}

\begin{deluxetable}{ccccc}
\tabletypesize{\scriptsize} 
\tablewidth{10cm}
\tablecaption{Photometry of nuclear region of NGC~5253}
\tablehead{\colhead{$\lambda$}	& \colhead{$f_\nu$}    
& \colhead{Aperture} &\colhead{Reference} &
\colhead{Uncertainty}\\
\colhead{$\mu$m} & \colhead{mJy}}
\startdata 
0.255 &	0.20 & 1.5"x1.6"      & WFPC2 F255W & 7\% \\
0.547 &	0.40 & 1.5"x1.6"      & WFPC2 F547M & 3\% \\
\smallskip
0.814 &	0.84 & 1.5"x1.6"      & WFPC2 F814W & 5\% \\
1.1   &	1.20 & 1.5"x1.6"      & NICMOS NIC2 F110W 	& 7\%\\
1.6   &	1.20 & 1.5"x1.6"      & NICMOS NIC2 F160W 	& 6\%\\
1.9   &	1.58 & 1.5"x1.6"      & NICMOS NIC2 F190N 	& 6\%\\
\smallskip
2.2   &	3.55 & 1.5"x1.6"      & NICMOS NIC2 F222M 	& 4\%\\
2.2   &	4.20 & 1.5"x1.6"      & Spex spectroscopy	& 10\%\\
3.0   &	19.0 & 1.5"x1.6"      & Spex spectroscopy	& 10\%\\
3.5   &	48.4 & 1.5"x1.6"      & Spex spectroscopy	& 10\%\\
\smallskip
4.0   &	92.5  & 1.5"x1.6"      & Spex spectroscopy	& 10\%\\
7.8   &	867.  & circular-8"   & Frogel et al. 1982   & 27\%\\		
8.6   &	817.  & circular-8"   & Frogel et al. 1982   & 13\%\\
9.6   &	1067. & circular-8"   & Frogel et al. 1982   & 14\%\\		
10.4  &	1541. & circular-8"   & Frogel et al. 1982   & 11\%\\		
11.4  &	1574. & circular-8"   & Frogel et al. 1982   & 12\%\\		
12.4  &	2041. & circular-8"   & Frogel et al. 1982   & 22\%\\		
10.0  &	1298. & circular-8"   & Frogel et al. 1982   & 6\%\\	
20.0  &	6098. & circular-8"   & Frogel et al. 1982   & 15\%\\
10.0  & 1500. & circular-5.4'' & Lebofsky \& Rieke 1979 & \\
\smallskip
21.0  & 2800. & circular-5.4'' & Lebofsky \& Rieke 1979 & 25\%\\
11.7  &	2200. & unresolved 	& Gorjian et al. 2001	 & 10\%\\
\smallskip
18.7  &	2900. & unresolved	& Gorjian et al. 2001	 & 10\%\\
4.5   &	143.  & point-source 	& ISOCAM &  4\%\\
6.7   &	520.  & point-source 	& ISOCAM &  9\%\\
7.7   &	829.  & point-source 	& ISOCAM &  7\%\\
12.0  &	2060. & $\simeq$point-source	& ISOCAM &  4\%\\
\smallskip
14.9  &	4100. & $\simeq$point-source 	& ISOCAM &  4\%\\
12.0$^*$  &	2730. & entire galaxy 	& IRAS-Roche et al. 1991  & \nodata\\
25.0$^*$  &	12630. & entire galaxy	& IRAS-Roche et al. 1991  & \nodata\\
60.0$^*$  &	30630. & entire galaxy	& IRAS-Roche et al. 1991  & \nodata\\
\smallskip
100.0$^*$ &	29490. & entire galaxy 	& IRAS-Roche et al. 1991  & \nodata\\
850.0$^*$ &  192.  & 41\arcsec      & SCUBA-James et al. 2002 & 12\%\\
\enddata
\tablecomments{$^*$Upper limits to the nuclear emission.\\
The errors account for the photometric calibration 
uncertainty, and when possible,  the uncertainty from 
estimating the contribution from the underlying
galaxy (see Section~5.1).
The absolute photometric calibration uncertainties for the ISOCAM data are
from the ISOCAM handbook (Blommaert et al. 2001).We also make use of the $8-13\,\mu$m spectrophotometry of 
Aitken et al. (1982).\\}
\smallskip
\end{deluxetable}

\begin{deluxetable}{cccc}
\small
\tablecaption{Extinction to the gas for the central region
of NGC~5253 for a foreground dust screen model.}
\tablehead{\colhead{Aperture}  & \colhead{Lines} &
\colhead{$A_V$} & \colhead{Ref.}}
\startdata
\multicolumn{4}{c}{Infrared estimates}\\
\hline
 $0.6\arcsec\times 3\arcsec$ & Br$\alpha$/Br$\gamma$ & $18\pm5$mag & 2\\
 $1.5\arcsec \times 1.6\arcsec$ &
Br$\alpha$/Br$\gamma$ & $13\pm2$mag & 1\\
& Pa$\alpha$/Br$\gamma$ & $13\pm5$mag & 1\\
& Br$\delta$/Br$\gamma$ & $6\pm5$mag & 1\\
 $10.3\arcsec \times 20.7\arcsec$ &
Br$\alpha$/Br$\gamma$ & $8\pm3$mag$^*$ & 3\\
Circular 5.4\arcsec & $9.7\,\mu$m Silicate feature &
10mag & 3,4\\
\hline
\multicolumn{4}{c}{Optical estimates}\\
\hline
Central $23\arcsec \times 23\arcsec$ & H$\alpha$/H$\beta$ 
&$1-3\,$mag$^{**}$ & 5\\
\enddata
\tablecomments{References: 1. This work. 2. Turner et al. 2003.
3. Kawara, Nishida, \& Phillips 
1989. 4. Aitken et al. 1982. 5. Walsh \& Roy 1989.\\
$^*$ $A_V$ obtained from the 
Br$\alpha$/Br$\gamma$ line ratio reported in Kawara et al. 
1989 and the extinction of Rieke \& Lebofsky 1985\\
$^{**}$ Observed range of $A_V$ in the
central region of NGC~5253.}
\end{deluxetable}

\clearpage

\section{The Nuclear Stellar Population}
\subsection{Age}
The spectroscopy through the $1.6\arcsec \times 
1.5\arcsec$ aperture can be used to derive the properties 
of the stellar population of the central cluster C1+C2, as
long as it is  not heavily contaminated by other faint
clusters/H\,{\sc ii} regions.
We can compare the Pa$\alpha$ luminosities of the H\,{\sc ii} regions
H1 and H2 (Section~3.2) believed 
to represent well the 
luminosities of the two nebulae, with that measured simulating  
the Spex $1.6\arcsec \times 1.5\arcsec$ 
aperture on the continuum subtracted 
Pa$\alpha$ image. We find that these
two measurements  are within $\simeq 15\%$ of each other, and thus 
we believe that the spectroscopic data 
are mainly sampling the double cluster/H\,{\sc ii} region.

The EWs of hydrogen recombination lines are reliable age estimators
of young (that is, the lines are in 
emission) stellar populations for the case of instantaneous
star formation (e.g., Alonso-Herrero et al. 1996) since 
EWs measure the ratio between massive (young) stars and the total number of
stars. Moreover, if the extinction to the gas is the same as the extinction
to the stars (cf. Calzetti 2001) 
then the EWs are extinction-independent age indicators.
The observed EW of Br$\gamma$
(Table~2) yields an age for the nuclear stellar population of C1+C2 of
approximately 
$3.4\pm0.2 \times10^6\,$yr, similar to that estimated by  
Davies, Sugai, \& Ward (1998) using
Br$\gamma$ narrow-band imaging. We have used Starburst99 
(Leitherer et al. 1999) 
for a stellar population formed in an  instantaneous burst 
with solar metallicity and a Salpeter IMF with lower and upper 
mass cutoffs of $M_{\rm low} = 1\,{\rm M}_\odot$ and 
$M_{\rm up} =100\,{\rm M}_\odot$, respectively.  This 
age estimate is not very dependent on the metallicity of the 
stars,  but it is rather sensitive to the choice of 
IMF (see Leitherer et al. 1999).

Using {\sc daophot} on the {\it HST}/NICMOS images we
have been able to separate both clusters/H\,{\sc ii}
regions and measure their near-infrared magnitudes and EW of Pa$\alpha$ 
(see Table~1).
Assuming that the age of young clusters is  
well represented by the observed EW of Pa$\alpha$, 
we find that C1 and C2 are both young star clusters,
with similar ages of approximately $3.3\pm1.0\times10^6\,$yr, in good 
agreement with other estimates. 
For instance, Calzetti et al. (1997) obtained an age 
of $2.2-2.8\times 10^6\,$yr for the peak of the H$\alpha$
emission (presumably our cluster C1) based on UV and optical photometric data,
and the EW of H$\alpha$. Tremonti et al. (2001) using UV spectroscopy
fitted the age of the brightest optical cluster to $2^{+0.7}_{-0.8}\times
10^6\,$yr.
Ground-based and {\it ISO} mid-infrared spectroscopy 
of the central region of NGC~5253 suggests a stellar population
$2-4 \times 10^6\,$yr old (Beck et al. 1996 and Crowther
et al. 1999).

The non-detection of the $K$-band CO bands (see Fig.~3)
sets an upper limit to the age of the population in the 
central double cluster of $7\times 10^6\,$yr.
We also extracted a circumnuclear $K$-band spectrum by subtracting
the $1.5\arcsec \times 1.6\arcsec$ spectrum from the 
$1.5\arcsec \times 3\arcsec$ one. Even in the region 
surrounding C1 and C2 there is no evidence
for the presence of CO bands. One possibility for the non-detection of 
the CO bands even in the circumnuclear region would be the presence 
of dust emission peaking in the mid-infrared. This component would fill in 
the CO bands in the $K$-band, and weaken them as observed in many Seyfert
1 and 2 galaxies with a strong thermal infrared continuum
(e.g., Ivanov et al. 2000). Indeed from Fig.~3 
it is clear that in the nuclear region the $2-4\,\mu$m continuum 
is very steep and may have a significant contribution from 
thermal emission (see discussion 
in Sections~5.2.2 and 7.2). This is not however the case for the circumnuclear 
spectrum where the continuum slope resembles 
that of less reddened starbursts (see Fig.~3). Thus we
believe that the lack of CO bands in the nuclear
region of NGC~5253 is not fundamentally due to the presence
of a red continuum. 

We note however that age estimates based on the CO bands may be dependent 
on the metal abundance ([Fe/H]) 
of the stars as there is a dependence of the strength of the near-infrared
CO bands on the metallicity of the stars (Origlia et al. 1997; 
Frogel et al. 2001; Ivanov et al. 2004).
The upper limit for the age 
of the stellar population based on the non-detection 
of the CO bands remains approximately constant ($7-9\times
10^6\,$yr) even if we consider the models with the lowest
metallicity in Starburst99 ($Z=\frac {1}{20}\,{\rm 
Z}_\odot$). This together with the measured EW of Br$\gamma$ 
of the circumnuclear region (Table~2) suggests that the stellar 
population within
the entire central $3\arcsec \times 1.6\arcsec$ 
($32\,{\rm pc} \times 60\,{\rm pc}$) region of NGC~5253 is very young 
($\simeq 3.5\times 10^6\,$yr), and
it is not only confined to the double star cluster C1+C2.

\subsection{Extinction}

The interpretation of the properties of the central double cluster
requires an accurate estimate of the 
extinction to the source. The extinction to the gas can be estimated 
from hydrogen recombination line ratios, so long as the comparison is 
based on measurements with the same beam sizes. To determine the 
extinction to the stars we can compare observed colors (or the 
SED) with the outputs of Starburst99 for
the age range derived in the preceding section. The determination
of the extinction to the gas and stars is described in the
next two sections.

\subsubsection{Extinction to the gas}
 
There have been a number of estimates of the extinction to the 
gas in the central
region of NGC~5253, based primarily on 
optical and near-infrared 
hydrogen recombination lines. The largest uncertainty in the 
extinction by far comes from the 
unknown distribution of dust within the source. Meurer et al. (1995) 
studied the  UV properties of a sample of starburst galaxies that 
included NGC~5253, and concluded 
that the dust geometry can be effectively described in terms of a 
foreground screen configuration 
near the starburst. However, their galaxies were 
UV-selected  and may be biased toward 
low extinction and possibly an unrepresentative distribution of the dust. 

We use the observed emission line fluxes
of Pa$\alpha$, Br$\delta$, Br$\gamma$ and Br$\alpha$ and the extinction
law of Rieke \& Lebofsky (1985) to infer the extinction to the 
central $1.5\arcsec \times 1.6\arcsec$. In Table~5 we show our results 
together with 
estimates from the literature using a foreground dust screen model. 
All the near-infrared
estimates of the visual extinction to the gas 
are consistent to within the errors, whereas the optical estimates 
are significantly lower (see discussion in Calzetti et al. 1997). 
The differing values of the extinction as a function 
of wavelength are the usual indication that the extinction is either 
patchy or mixed with the emitting sources along the line of sight (or both). 
Indeed,  Calzetti et al. (1997) have 
argued that a young star cluster will be 
deeply embedded in its natal molecular cloud and thus the stars, gas and 
dust are likely to be mixed, providing an efficient way of ``hiding'' 
the dust. 

Using a value of the extinction of $A_V = 11\,$mag (which is
consistent with all our estimates) and the observed
fluxes through the $3\arcsec \times 1.5\arcsec$ apertures we
infer a number of ionizing photons $N_{\rm Ly}= 4 \times 10^{52}\,
{\rm s}^{-1}$. This is similar to, although slightly lower than, 
the radio estimates of Turner et al. 
(2000, 2003) and Turner \& Beck (2004).
The radio H92$\alpha$  observations of Mohan et al. (2001) predict 
$N_{\rm Ly} \simeq 2-4 \times 10^{52}\,
{\rm s}^{-1}$, but this line is only sensitive to a narrow
range of nebular densities.

\subsubsection{Extinction to the stars and dust emission at 
near-infrared wavelengths}

A qualitative description of the extinction to the 
stars can be obtained from color maps (see the 
$J-H$ color map of the central $12\arcsec \times 12\arcsec$ in
Fig.~2) and false color images (e.g., Fig.~5, and also
Turner et al. 2003). The C2 cluster is heavily obscured, but it becomes
the brightest source in this galaxy at $2.2\,\mu$m. 
On the other hand, the region surrounding C2 and the region 
to the southeast of C2 appear blue with respect to other star clusters due 
to the youth of the ionizing stars. 
Also, there is evidence for the dust lane that crosses 
the galaxy approximately
east-west.  

Evidently the high value of the central
extinction derived in the previous section
must be mostly associated with C2, as the infrared colors of C1 appear
to be quite blue (see Table~1) consistent with outputs from Starburst99
for a young starburst. A 
stellar population created in an instantaneous burst after 
$\simeq 3-3.4\times 10^6\,$yr shows infrared colors: 
$J-H=0.12-0.04$ and $H-K=0.35-0.19$. The observed near-infrared 
colors of C1 are consistent with $A_V \simeq 2\pm 2\,$mag. The 
$J-H$ color of C2 is consistent with $A_V \simeq 14\pm 2\,$mag
for the same age range as above. Using a model where the central 
cluster is hidden behind a thin dust layer
and is completely embedded in a thick dust cloud Calzetti et al.
(1997) found that values of the extinction $A_V > 9\,$mag were 
necessary to account for the observed UV-optical colors of the central 
cluster. The dereddened $H-K$ color of C2 is still very red. As we
shall see in Section~7, this is 
the effect of hot dust emission with a significant
contribution in the $K$-band and/or differential 
extinction. If we assume that there is no differential extinction
between the $H$ and the $K$-band, then approximately 70\% of the 
$K$-band emission in C2 could be produced by hot dust (with some small 
contribution from gas emission as well).

\subsection{Stellar mass of the double cluster}

Using the age and the extinction estimates, 
we can infer the stellar masses of the two clusters for a Salpeter IMF with 
mass cutoffs of $M_{\rm low} = 1\,{\rm M}_\odot$ and 
$M_{\rm up} =100\,{\rm M}_\odot$, using their 
absolute $H$-band magnitudes. For C1 we estimate
a stellar mass of $M_{\rm C1} \simeq 5 \times 10^4\,
{\rm M}_\odot$ (see also Fig.~6). If we continued the high mass slope
all the way down to $0.1\,{\rm M}_\odot$ the  total 
stellar mass would be $M_{\rm C1} 
\simeq 1.3 \times 10^5\,{\rm M}_\odot$. For the C2 cluster since the current
dataset does not allow us to distinguish between  hot dust emission 
in the $K$-band and differential extinction, we can only obtain a 
range  of stellar masses. If there is no differential 
extinction between the gas and the stars, and 
up to 70\% of the $K$-band emission is due to 
hot dust,  then the stellar 
mass of C2 would be $M_{\rm C2} > 3 \times 10^5\,
{\rm M}_\odot$ (or $M_{\rm C2} > 7.7 \times 10^5\,
{\rm M}_\odot$ for a Salpeter IMF extending down 
to $0.1\,{\rm M}_\odot$). 
The upper limit to this estimate would be 
if all the $K$-band emission were stellar in origin (implying that 
there is differencial extinction) and 
we corrected the observed $H-K$ color to that corresponding
to $3-3.4$\,Myr; in this 
case the stellar mass would be $M_{\rm C2} < 1 \times 10^6\,
{\rm M}_\odot$ (or $M_{\rm C2} < 2.6 \times 10^6\,
{\rm M}_\odot$ for an IMF down to $0.1\,{\rm M}_\odot$). 
In any case, C2 dominates C1 by mass for any
plausible C2 age and extinction. Using the number of
ionizing photons (corrected for extinction)
 derived in Section~5.2.1 we would require 
a mass for the two ionizing clusters of  
$M_{\rm C1+C2} \simeq 1 \times 10^6\,
{\rm M}_\odot$ (or  $M_{\rm C1+C2} = 2.6 \times 10^6\,
{\rm M}_\odot$ for an IMF down 
to $0.1\,{\rm M}_\odot$), for the same star formation history.

\section{$H$-band selected star clusters in the central $\simeq 20\arcsec$}

\subsection{Bright clusters}
In Table~1 we give the photometry, colors and 
observed EW of Pa$\alpha$ of bright clusters ($M_H \le -11\,$mag) 
selected in the $H$-band. In the last column of this table 
we give possible associations with star clusters detected at
other wavelengths. The correspondences are solely based on 
the measured relative offsets taking into 
account the fact that C1 is easily identified in the UV and 
optical, and at near-infrared wavelengths $\lambda < 2\,\mu$m. 

We find near-infrared counterparts to five of the
six star clusters optically selected by Calzetti et al. (1997), as well
as some of the clusters reported by Harris et al. (2004), 
as shown in Table~1. The absolute $M_{\rm F160W}$ and $M_{\rm F222M}$ 
magnitudes (not corrected for extinction) are typical of near-infrared star 
clusters detected in other hot-spot galaxies and starbursts 
(e.g., Alonso-Herrero, Ryder, \& Knapen 2001; 
Elmegreen et al. 2002; Maoz et al. 2001). Except
for star cluster C2, the rest of the 
central clusters in NGC~5253  are not as bright as the near-infrared
super star clusters identified in luminous and ultraluminous infrared 
galaxies (Scoville et al. 2000, Alonso-Herrero et al. 2002) 
and interacting galaxies (e.g., the Antennae, Gilbert et al. 
2000; Mengel et al.
2002).

At ground-based resolution  the star cluster C3+C3' appears 
as the second brightest near-infrared
cluster after C1+C2, and was identified 
by Davies, Sugai, \& Ward (1998) as one of the bright
three near-infrared hot 
spots (see Figs.~1 and 2) in this galaxy. The third brightest cluster 
identified in Davies et al. (1998) corresponds to 
the C4+C5 cluster in our notation (see Table~1).
The $K$-band spectrum of 
C3+C3' is presented in Fig.~3. Note that the 
$1.5\arcsec \times 1.6\arcsec$ extraction aperture possibly 
includes other fainter star clusters (see e.g., Fig.~2). The 
measured EW of Br$\gamma$ sets a
lower limit to the age of this cluster of approximately $>7\times 10^6\,$yr, 
using outputs of Starburst99 under the same assumptions
used for the central cluster. This age is consistent with that
derived from the observed EW of Pa$\alpha$ ($7.5\pm0.5\times10^6\,$yr). 
The lack of $2.3\,\mu$m CO bands, on the 
other hand, sets an upper limit to the age of the cluster 
of approximately $7-9$ million years, although this limit is 
dependent on the metallicity of the stars.
Davies et al. (1998) using the measured EW of Br$\gamma$ 
($=14-16\,$\AA; note that they used a 2\arcsec-diameter aperture) 
from narrow-band
imaging derived an age for this cluster of 
10 million years, although using Starburst99 we 
would get $\simeq 6.5\times 10^6\,$yr. Calzetti et al. (1997) 
estimated an age for C3 of $8-12\times 10^6\,$yr.

Tremonti et al. 
(2001) studied in detail clusters C4 and C5 (clusters 
NGC5253-3 and NGC5253-2 in their notation) using
{\it HST} UV spectroscopy,
and derived ages of 3\,Myr and 8\,Myr, respectively. Their
result is in good agreement with the age from 
the measured EW of Pa$\alpha$
of C5 (see also Harris et al. 2004). However the young age 
Tremonti et al. (2001) inferred 
for C4 appears to be 
inconsistent with the lack of bright Pa$\alpha$ emission 
at the location of this star cluster. Harris et al. (2004) on 
the other hand find an age for this cluster of between 10 and 14\,Myr 
consistent with our finding. For other near-infrared clusters in
common with those optically selected by Harris et al. (2004)
we find ages and stellar masses (see next section) consistent with   their 
photometric and EW of
H$\alpha$ based estimates.

\clearpage

\begin{figure*}
\figurenum{6}
\epsscale{0.9}
\plotone{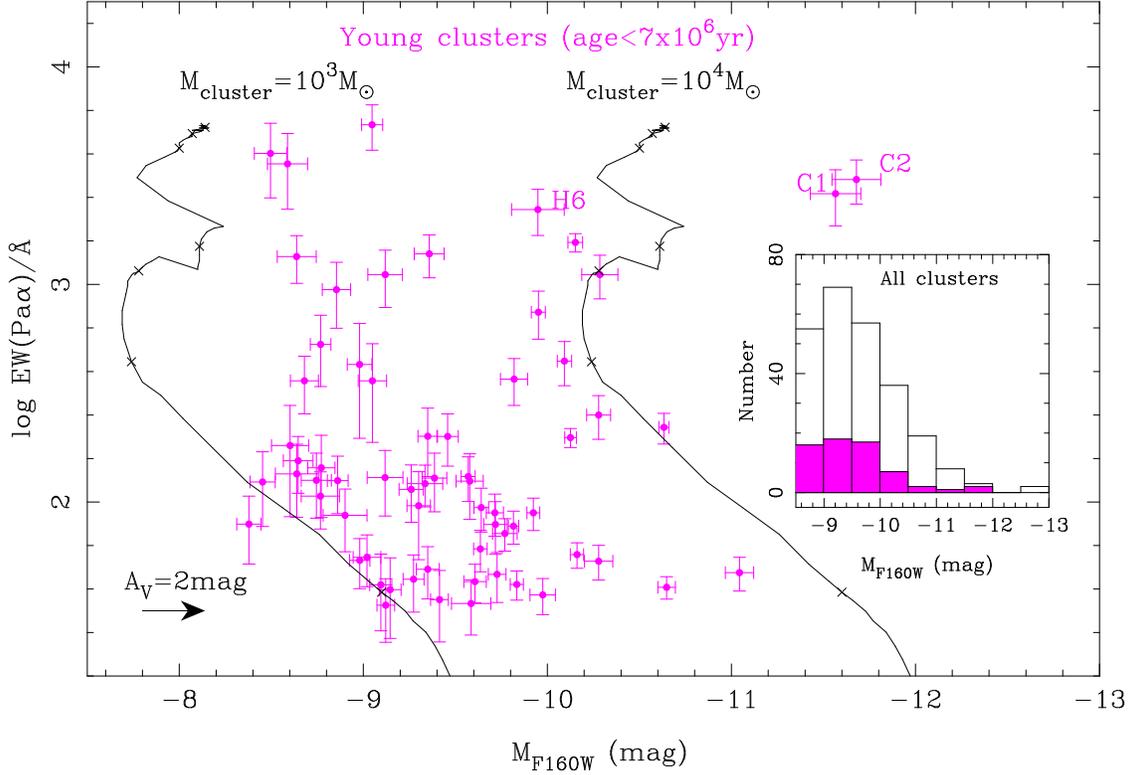}
\caption{{\it Main Panel:} --- 
Absolute $M_{\rm F160W}$ magnitude vs. equivalent width of
Pa$\alpha$ emission line for $H$-band 
selected clusters (filled dots).
We plot young star clusters with observed 
$\log {\rm  EW(Pa}\alpha)/{\rm \AA} \ge 1.5\,$, that is, star clusters 
younger than approximately $7\times 10^6\,$yr. We mark
the positions of the central double cluster C1 and C2 (see
Section~4.2) and H\,{\sc ii} region H6 (see Section~6.2).
The magnitudes have not been corrected for extinction.
The arrow shows what would be the effect of correcting the observed
absolute $M_{\rm F160W}$ for 2 magnitudes of visual extinction
assuming the same extinction to the stars and the gas.
The lines represent the time evolution of 
star clusters with 
masses $M=10^3\,{\rm M}_\odot$ and $M=10^4\,{\rm M}_\odot$ 
for instantaneous star formation, a Salpeter IMF 
($M_{\rm low} = 1\,{\rm M}_\odot$ and 
$M_{\rm up} =100\,{\rm M}_\odot$), 
and solar metallicity using Starburst99 (Leitherer et al. 
1999). The crosses on 
the model lines are drawn at intervals of 1 million years,
with the youngest age on the upper part of the curves.
{\it Insert:} --- The empty histogram shows the distribution of the 
absolute $M_{\rm F160W}$ magnitudes of all the star clusters selected 
in the $H$-band, whereas the filled histogram shows the clusters with ages 
younger than 7 million years, that is, those shown in the main panel.}
\end{figure*}

\begin{figure*}
\figurenum{7}
\hspace{2cm}
\psfig{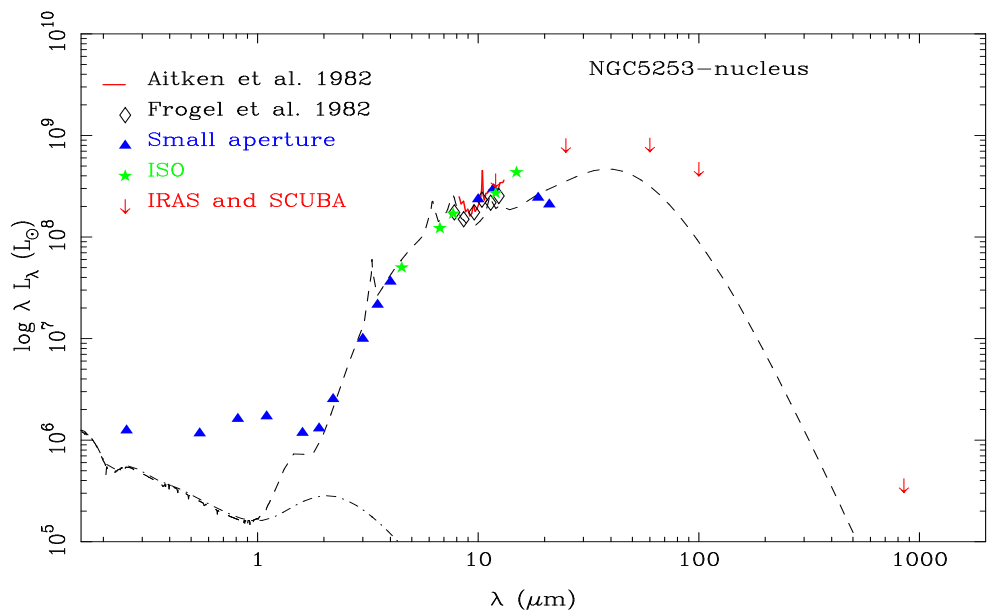}
\caption{SED of the nuclear emission of NGC~5253. The small aperture 
measurements are from this work, Gorjian et al. (2001) and 
Lebofsky \& Rieke (1979). As discussed in Section~3.3 the $20\,\mu$m
flux from Frogel et al. (1982) is not used. 
For the nuclear SED the IRAS and Scuba 
data points are plotted as upper limits because these measurements
comprise the entire galaxy. The best fit to the nuclear SED of
NGC~5253 is the dashed line, whereas the dashed-dotted line
shows the stellar contribution to this fit. For details on the modelling 
see Section~7.}
\end{figure*}

\clearpage

\subsection{Young versus old star clusters}
The only optically bright star cluster that is not
bright in the near-IR is NGC5253-4,  using the notation 
of Calzetti et al.
(1997). This star cluster, however, is detected  
as a bright H\,{\sc ii} region (H6, see Table~3) and is also 
the brightest UV source (Meurer et al. 1995). 
Based on the observed EW of Pa$\alpha$ we derive a very
young age for this cluster/H\,{\sc ii} region 
($\simeq 3.5\times 10^6\,$yr), similar to that of C1 and C2. 
Harris et al. (2004) derive an age for H6 (their cluster \# 4) of 
$5\times 10^6\,$yr using the EW of H$\alpha$. The youth of H6
explains why the ionizing cluster has not become infrared 
bright (that is, its absolute magnitude is 
fainter tha $M_{\rm F160W} = -11\,$mag).
Starburst99 suggests that if the H\,{\sc ii} region H6  was formed in  
an instantaneous 
burst the ionizing cluster will become near-infrared
bright in the next three or four million years (see Fig.~6 for 
the location of H6 and future time evolution 
of its near-infrared magnitude). The age derived from the EW of Pa$\alpha$  is consistent
with that derived in Calzetti et al. (1997) using UV and 
optical magnitudes, together with the EW of H$\alpha$. 

As we have shown, the EW of Pa$\alpha$ is a useful indicator of the 
age of young star clusters\footnote{However, 
in general for higher metallicity galaxies, Pa$\alpha$
must be used in this way with some caution due to the
delayed episode of ionization from WR stars (Rigby \& Rieke 2004).}. 
For all the 269 clusters detected in the $H$-band
we can determine from a statistical point of view both the masses of
the young clusters, and what fraction of all the detected infrared 
clusters are young.
The main panel of Fig.~6 shows the  absolute $M_{\rm F160W}$ magnitude 
vs. the EW of Pa$\alpha$ for young clusters selected in 
the $H$-band. We show clusters with observed EW of Pa$\alpha \ge 32\,$\AA \
or ages of less than $\simeq 7\,$million years.
We also show outputs of Starburst99 (Leitherer et al. 1999) for clusters 
with masses $M_{\rm cluster}=10^4\,{\rm M}_\odot$ and 
$M_{\rm cluster}=10^3\,{\rm M}_\odot$ formed 
in an instantaneous burst with a Salpeter IMF between 
1 and $100\,{\rm M}_\odot$. The arrow shows what would be 
the effect of correcting the observed
absolute $M_{\rm F160W}$ magnitudes by  
2\,magnitudes of visual extinction for the case of 
a foreground dust screen (assuming 
extinction to stars and line emission is the same). From the 
colors of bright clusters (Table~1) and the color map in Fig.~2 it is 
clear that the young clusters are not very obscured, and thus the 
observed absolute  magnitudes can be used to infer
the stellar mass. From Fig.~6 we find that the young clusters have 
stellar masses of between $10^3\,{\rm M}_\odot$ and 
$10^4\,{\rm M}_\odot$, in good agreement with the findings of 
Harris et al. (2004). Unfortunately for older clusters where we do not
have EW of Pa$\alpha$ based ages,  
we cannot derive the stellar masses as the 
near-infrared mass-to-light ratio is dependent on the age of the 
stellar population.

The insert of Fig.~6 also shows the distribution of absolute
$M_{\rm F160W}$ magnitudes of all the clusters detected and that of the 
the young (age  $7$ million years) clusters. 
We find that approximately $20-30$\% of the clusters detected in 
the $H$-band have ages of less
than 7 million years. This fraction remains roughly constant 
in all magnitude bins.

\clearpage

\begin{deluxetable}{ccc}
\tablecaption{Results from the modelling of the SEDs of the nuclear
region and galaxy.}
\tablehead{\colhead{Property} & \colhead{Nuclear Region} & \colhead{Entire Galaxy$^1$}}
\startdata
Star Formation & Burst & Infall model \\
Age of stellar population & $t= 3.2 \times 10^6\,{\rm yr}$ & $t= 30\times 10^6\,$yr\\
Effective optical depth & $\tau_V({\rm eff}) = 15.7$ & 
$\tau_V({\rm eff})=0.9$\\
Size of the emitting region & $r_t=21\,$pc & $r_t=252\,$pc \\
Mass of HI & $M{\rm (H)} = 6.2 \times 10^6\,{\rm M}_\odot$ &
$M{\rm (H)} = 5.0 \times 10^7\,{\rm M}_\odot$ \\
Dust Mass &   $M_{\rm d} = 5.0 \times 10^4\,{\rm M}_\odot$&
$M_{\rm d} = 9.5 \times 10^4\,{\rm M}_\odot$ \\
Stellar Mass$^2$ &  $M_*  = 2.8 \times  10^6\,{\rm M}_\odot$ &
$M_*  = 1.0 \times  10^7\,{\rm M}_\odot$ \\
Bolometric Luminosity &  $L_{\rm bol} = 1.0 \times 10^9\,{\rm L}_\odot$ 
& $L_{\rm bol} = 1.8 \times 10^9\,{\rm L}_\odot$ \\
\enddata
\tablecomments{$^1$Results from the modelling of 
the entire galaxy are from Takagi et al.
(2003b), using an infall model of 
chemical evolution with a timescale of 
$t_0=100\,$Myr  (i.e., SFR $\propto M_{\rm gas}/
t_0$), recomputed
for the distance assumed in this paper ($d=4.1\,$Mpc).\\
$^2$Stellar masses are for a Salpeter IMF with mass cutoffs of
$M_{\rm low} = 0.1\,{\rm M}_\odot$ and 
$M_{\rm up} =60\,{\rm M}_\odot$.}
\end{deluxetable}

\clearpage

\begin{figure*}
\figurenum{8}
\hspace{1cm}
\psfig{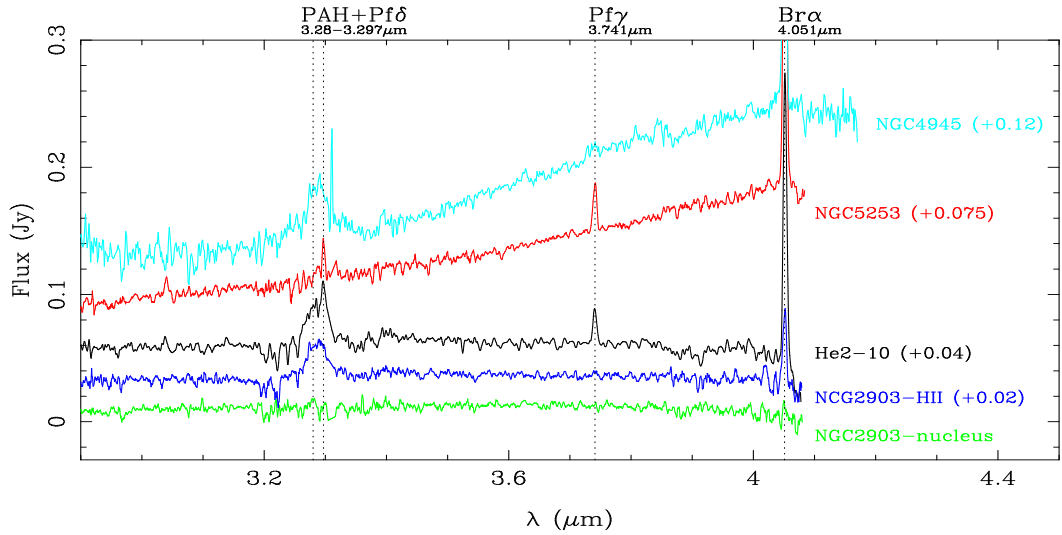}

\caption{Spex $L$-band spectrum of NGC~5253 and 
comparison galaxies, 
except NGC~4945 for which the data are from Spoon et al. (2003). The spectra 
have been shifted to rest-frame wavelengths. We also mark the most prominent 
emission features.}
\end{figure*} 

\clearpage

\section{The Nuclear SED of NGC~5253}

\subsection{The Model}

We used  the SED model of Takagi, Vansevicius \& Arimoto (2003) to reproduce
the UV to mid-infrared SED of the nuclear region of NGC~5253
(Table~4). This model makes
use of  the population synthesis model of Kodama \& Arimoto (1997) to
predict the stellar emission. For the nucleus of this galaxy we assumed 
an instantaneous burst of star 
formation with  solar metallicity and a Salpeter IMF 
with lower and upper mass cutoffs of $M_{\rm low} = 0.1\,{\rm M}_\odot$ and 
$M_{\rm up} =60\,{\rm M}_\odot$, respectively. In the SED model of 
Takagi et al. (2003a), stars are centrally concentrated in a spherical 
region with a density profile given by the King model, 
while the dust grains are homogeneously distributed throughout the 
emitting region. 
The equations of radiative transfer are numerically solved by 
considering multiple scattering of photons and self-absorption by dust
(see Takagi et al. 2003a for more details). The main  
parameters of this model are the age of the stellar population, 
the equivalent $V$-band optical depth $\tau_V$ 
measured from the center to the outer edge of the region, 
and the size of the emitting region, $r_t$. 
Note that $\tau_V$ is defined by the column
density of dust, whereas $\tau_V({\rm eff})$ is defined by the luminosity
ratio of input and output at $V$-band. 
The latter optical depth was 
used to verify the SED models of star forming galaxies of 
Takagi, Arimoto, \& Hanami (2003b) because it is
directly comparable with observations.
The extinction curve can be chosen from 
three types: MW, LMC, and SMC. 
Using this model, we manually found the best-fitting model to 
the nuclear SED, i.e. we calculated a new model in each step 
of the fitting process until convergence was reached. 

\subsection{Results from the modelling}
In Table~6 the results of the modelling of the nuclear 
SED of NGC~5253 are presented
and compared with those from the modelling of the 
entire galaxy emission (Takagi et al. 2003b).
The fit to the nuclear SED is shown in Fig.~7.
There is a degeneracy in several of the model parameters --- 
age, $\tau_V$, and $r_t$ ---  for
the overall nuclear SED from $\lambda \simeq 1\,\mu$m to 
$\simeq~10\,\mu$m, but the depth of tbe silicate feature at $9.7\,\mu$m
can be used to constrain the model parameter ranges. 
We could not obtain reasonable fits 
with either a LMC or  a SMC-type extinction curve. For both
these extinction curves the fraction 
of silicate grains is larger than that in the MW-type 
(see Takagi et al. 2003a). We found a conservative limit of 
$\tau_V < 30\,$ from the observed 
silicate absorption (see also Aitken et al. 1982; Kawara et al. 1989
and Fig.~7) 
with the MW-type extinction curve. 
With this $\tau_V$, the observed nuclear SED from $\sim 1\,\mu$m to 
$10\,\mu$m was reproduced with SEDs in which the contribution 
from red supergiants at near-infrared wavelengths is still negligible, 
i.e. $< 4$ Myr in the adopted population synthesis model.
An older age would need values of the optical depth 
$\tau_V > 30\,$ to fit the SED from 
$\simeq 1\,\mu$m to $10\,\mu$m because the relative brightness of the 
intrinsic SED in the near-infrared increases
as the system ages. Since the observed depth of 
the silicate feature is relatively 
shallow, we fit values of the optical depth $\tau_V \le 20\,$  and 
the age $t < 3.5\,$Myr. The fitted age is in good  
agreement with the values obtained in Section~5.1. 
To compare the fitted optical depth with observations of the 
gas emission, we need to use $\tau_V({\rm eff})$. 
When  $\tau_V$ is large, 
it is not straightforward to compare $\tau_V({\rm eff})$ with 
$A_V$ as derived from the emission lines.
For a meaningful comparison, we need to choose
a wavelength to calculate $\tau({\rm eff})$ at which there be 
still significant stellar emission from 
the central region and be close to the wavelengths used to estimate
$A_V$. For the model with $t=3.2\,$Myr and $\tau_V =20$, 
we derived an effective optical depth at $2.2\,\mu$m of 
$\tau_K({\rm eff})= 1.4$ from the luminosity ratio of the input and 
the attenuated stellar emission. For the MW extinction curve, 
this corresponds to $\tau_V({\rm eff})$ = 15.7 or 
$A_V= 17\,$mag. This is in good agreement with the 
estimate of $A_V$ from the infrared lines in Section~5.2.1.

Considering these results, the best fit to the nuclear SED is reached for a 
stellar mass of  $2.8\times 10^6\,{\rm M}_\odot$ (for 
a Salpeter IMF with mass cutoffs
$M_{\rm low} = 0.1\,{\rm M}_\odot$ and 
$M_{\rm up} =60\,{\rm M}_\odot$) formed in an instantaneous
burst,  with an age of 3.2\,Myr, $\tau_V = 20\,$ with the MW 
extinction curve. The fitted size for the emitting region is 21\,pc. 
The stellar mass needed to account 
for the observed nuclear SED of NGC~5253 is in good agreement with 
the stellar mass of the double cluster C1+C2 derived in Section~5.3
(taking into account the correction factor to use the same lower
mass cutoff of the IMF). The gas and dust masses are
$6 \times 10^6\,$M$_\odot$  and $5\times 10^4\,$M$_\odot$, 
respectively. The resulting bolometric luminosity is 
$10^9\,$L$_\odot$. One interesting result from comparing the 
fitting results of the nuclear and galaxy SEDs (Table~6) is that up to 
half of the total dust mass in NGC~5253 appears to be 
concentrated in the nuclear 
region, with the consequence that a similar fraction 
of the bolometric luminosity is due to the central cluster.

We find that the fluxes at $\lambda < 1\,\mu$m are  
not well reproduced with this model (Fig.~7). It is possible 
that these fluxes are still contaminated  by background radiation 
from the underlying galaxy (that is, not arising from 
the double cluster) and/or photons leaking 
through gas clouds, an effect that is not taken into account in the 
model. However a more likely explanation is the complicated geometry
of the central double cluster. As discussed in Section~4, most of the 
UV and optical fluxes measured through the small 
aperture come from cluster C1 which is not significantly obscured, 
at least as infered from its observed 
near-infrared colors (Table~1 and Section~5.2.2) and color maps (Fig.~2). 
Cluster C2 on the other hand, only starts becoming visible in the near-infrared
and dominates the nuclear 
emission at wavelengths longer than approximately 
$2\,\mu$m, indicating that it is deeply embedded. It is clear that 
the near-infrared to mid-infrared SED is dominated by C2, and thus the 
UV and optical fluxes should be taken as upper limits to the emission of
C2. This is suggestive of different dust geometry and optical depth 
for each of the two nuclear clusters. 

The ISO $15\,\mu$m flux and the ground-based $20\,\mu$m flux observed 
with the large aperture size (Frogel et al. 1982) could not be well 
reproduced with the model, indicating
that there may be some contribution from emitting sources other 
than the central double cluster. Note however, that 
the smaller aperture fluxes of Lebofsky \& Rieke (1979) 
and Gorjian et al. (2001) are well reproduced with the model. 
At $\lambda < 20\,\mu$m, the emission from very small 
grains (graphite) dominates,
and because of this, there is a 'shoulder' in the SED. 
Although the small aperture
$20\,\mu$m data seems to indicate a single peak of hot 
dust emission,  no SED model
has such a single peak. If we reduced the size of the 
emitting region to move the peak of the SED toward
shorter wavelengths, then the
near-infrared fluxes would be significantly over-estimated. 
Therefore, there should be a true
peak at $30\,\mu$m due to large grains, unless the emission 
from very small grains is  significantly deficient.
Finally, the best fit model shows that the $3.3\,\mu$m 
polycyclic aromatic hydrocarbon  (PAH) feature 
is still present, but is not very prominent since the 
continuum produced by very small grains is high, effectively reducing the 
equivalent width of this feature. 
A similar fit to the nuclear SED of NGC~5253 could be obtained if 
we reduced the amount of PAH carrier, since the infrared SED
remains the same, except for the PAH feature (i.e., PAHs have negligible
contribution to continuum). We return to this issue in Section~7.3.

\subsection{The non-detection of the $3.3\,\mu$m PAH feature 
in the nucleus of NGC~5253: age or
metallicity effect?}

Most massive star-forming galaxies display infrared emission features
at 3.3, 6.2, 7.7, 8.6, and $11.3\,\mu$m (Roche et al.\ 1991;
Uchida et al.\ 2000; Helou et al. \ 2000) that have 
been attributed to bending and
stretching modes of CH and CC bonds in 
PAHs (Duley \&
Williams 1981; L\'eger \& Puget 1984; Puget \& L\'eger 1989).  
Among others, Roche et al.\ (1991) and Clavel et al.
\ (2000) have shown that the 
strength of the $11.3\,\mu$m PAH feature depends
strongly on environment: galaxies classified as Seyfert 1 show little to
no $11.3\,\mu$m emission, while many type 2 AGN 
(see also Imanishi, 2002 and  2003 for a study of the 
$3.3\,\mu$m feature) and most pure starburst 
galaxies show strong PAH 
features.  This behavior implies that as the radiation field density
increases, the PAH carrier is progressively destroyed.  This picture
is supported by M82 where the $3.3\,\mu$m PAH 
emission is found to be  anti-correlated with both 
the centrally peaked continuum emission and also
with the Br$\gamma$ emission (Normand et al. 1995).  A similar 
situation is observed in NGC~253 where there is a decrease
of the ratio of $3.3\,\mu$m PAH emission to continuum at the 
sites of strong star formation (Tacconi-Garman et al. 2004).

Dwarf galaxies in the metallicity range $12+\log{\rm O/H}=7.6-8.6$  
tend to show weak or no mid-infrared PAH features from ISO data 
(Madden 2002;  Thuan, Sauvage, \& Madden 1999) or ground-based observations 
(Roche et al. 1991), including NGC~5253 with 
$12+\log{\rm O/H}=8.4$. 
We are obtaining $L$-band spectroscopy of a sample of 
starburst galaxies with a range of metallicities using Spex on the IRTF.
Fig.~8  are the spectra for some of these galaxies, including  
a second dwarf galaxy, He2-10, with an age of $\simeq 5-6\,$Myr and 
metallicity of
$12+\log{\rm O/H}=8.9$ (Kobulnicky, Kennicutt, \& Pizagno 1999). 
For NGC~2903 we have spectra 
for the quiescent nucleus and one of the bright H\,{\sc ii} regions 
to the north of the nucleus (see Alonso-Herrero et al. 
2001). The $3.3\,\mu$m feature is
clearly detected in He2-10 and in the bright H\,{\sc ii} region of
NGC~2903, as well as in all 
starburst galaxies in our sample with 
metallicities $12+\log{\rm O/H}>8.9$.
We also show that the  highly obscured starburst/AGN 
NGC~4945 ($L$-band spectroscopy from Spoon et al. 2003) has bright 
$3.3\,\mu$m PAH emission, indicating that this feature can be 
detected regardless of the  extinction (Imanishi 2002).

The effects of metallicity and a harsh environment are coupled 
for young starbursts as in the nucleus of NGC~5253. As metallicity
decreases stars present harder spectra, and Wolf-Rayet stars require
larger mass progenitors thus creating a hard and intense 
radiation field that 
could suppress the formation of the $3.3\,\mu$m PAH feature. 
As discussed in Tacconi-Garman et al. (2004) and references therein, 
 in strong radiation fields 
the efficiency of $3.3\,\mu$m PAH emission can be lowered by 
photoionization of the PAH molecules, or by destruction of the PAH molecules.  
Rigby \& Rieke (2004) have demonstrated that 
the observed mid-infrared fine structure lines of the low metallicity 
starbursts
NGC~5253 and II~Zw40 can only be reproduced 
with a Salpeter IMF extending to $40-60\,{\rm M}_\odot$, whereas for their
sample of high mass, solar metallicity starbursts 
the interstellar radiation field will  not include the outputs of stars
above $40\,{\rm M}_\odot$ because most of them are probably embedded in
compact H\,{\sc ii} regions.
Although the number of low metallicity starbursts analyzed 
in Rigby \& Rieke (2004) is small, there seems to be an indication that 
in these  systems the interstellar medium is less effective in 
confining ultracompact H\,{\sc ii} regions,  and thus a harsher 
radiation field environment is present that can effectively 
decrease or even suppress 
the $3.3\,\mu$m PAH emission.

\section{Discussion and Conclusions}

We have analyzed {\it HST}/NICMOS observations ($1.1-2.2\,\mu$m)  and 
obtained $1.9-4.1\,\mu$m spectroscopy of the central region of the dwarf
galaxy NGC~5253. The NICMOS images have revealed the presence of 
a double cluster (C1+C2) in the nucleus of the galaxy 
separated by $0.3-0.4$\arcsec \ or 
$6-8$\,pc for the assumed distance of $d=4.1\,$Mpc.
This double cluster is also a bright double source of Pa$\alpha$ emission. We 
have also analyzed {\it HST}/WFPC2 observations and found that 
the western cluster (C2) is almost entirely obscured at UV and optical 
wavelengths, but becomes the brightest source in the galaxy at
$\lambda > 2\,\mu$m. The double cluster C1+C2 appears to be coincident
with the double radio nebula detected at 1.3\,cm by Turner et al. (2000), and
it is likely to be responsible for most of the mid-infrared nuclear 
emission in this galaxy (see also Turner \& Beck 2004).

The high spatial resolution of the {\it HST}/NICMOS images allows us 
to measure the near-infrared magnitudes, EW of Pa$\alpha$ and Pa$\alpha$
luminosities of C1 and C2. If the star formation in the double
cluster C1+C2 occurred in an instantaneous burst, C1 and C2 
have virtually the same age ($3-4\,$ million years). This young age for
the double cluster is consistent with estimates by other authors for 
cluster C1 from UV/optical imaging and spectroscopy (e.g., Calzetti 
et al. 1997; Harris et al. 2004), and the presence of Wolf-Rayet features
in the optical spectrum of the nucleus of the galaxy (Walsh \&
Roy 1987; Conti 1991). Although
both nuclear clusters appear to be coeval, C2 is more massive and 
obscured than C1. The stellar mass for C2 is in the range of
$M_{\rm C2} = 7.7 \times 10^5 - 2.6 \times 10^6\,{\rm M}_\odot$ 
(for a Salpeter IMF in the mass range 
$0.1-100\,{\rm M}_\odot$), depending on the dust emission contribution to the 
observed $K$-band magnitude, thus putting C2 in the category of 
super star clusters. The stellar mass of C1 is 
$M_{\rm C1} \simeq 1.3 \times 10^5\,{\rm M}_\odot$. 

The fact that both C1 and C2 appear to be coeval but C2 is much more
massive than C1 appears to conflict with the naive expectation 
that more massive clusters would blow away their birth clouds first.  
Using $K$ and $L$-band hydrogen recombination lines and near-infrared 
colors we have inferred a very high obscuration for C2,  suggesting 
that this young star cluster may be still (partially) embedded in its
 natal cloud. The age derived using the EW of 
Pa$\alpha$ assumes that 
the extinction to the stars and gas is the same. If  this is not
the case, as discussed in Calzetti et al. (1997), that is, 
stars, gas and dust are mixed, then the derived 
age of C2 would be only an upper limit. 

Turner \& Beck (2004) have put forward some alternative explanations to
that of the presence of a coeval double cluster in the nucleus of
NGC~5253. One possibility they discuss is that 
cluster C1 is not a cluster
but a reflection nebula from a gap opening in the cocoon around
the infrared cluster. Their second 
scenario is that the supernebula is cometary and that what 
we are seeing in the nucleus of NGC~5253 is motion of the H\,{\sc ii}
region/cluster to the west and north. We favor the interpretation of
a double cluster in the nucleus of NGC~5253 
that has resulted from triggered star formation, due to infalling gas along 
the minor axis of the galaxy (see Meier et al. 2002). 
The south-east elongation of 
the radio supernebula detected by Turner \& Beck (2004) shows 
a morphology similar to the Pa$\alpha$ emission (Fig.~4, left panel), 
but we find that the bright nuclear Pa$\alpha$ emission is extended over 
scales of a few arcseconds, whereas the radio supernebula is very
compact ($\simeq 0.05\arcsec$, see Turner \& Beck for more
details). From the $H$-band NICMOS image of the central $3\arcsec \times
3\arcsec$ it is clear that there are a number of near-infrared clusters with
no Pa$\alpha$ emission, but there must be other ionizing clusters
(not detected as bright sources of  $1.6\,\mu$m emission) that are responsible
for the extended Pa$\alpha$ emission. Moreover, we have 
derived a very young age ($3-4$ million years) using 
the $K$-band spectrum for the circumnuclear
region {\it excluding} C1 and C2. 
This all seems to indicate that what we are seeing are the 
effects of self-propagating star formation  
in the central region of NGC~5253.

In addition to the nuclear double cluster, we 
have identified a total of 269 star clusters in the $H$-band over 
the observed region of $19\arcsec \times 19\arcsec$, and 
measured their near-infrared magnitudes as well as the EW of Pa$\alpha$. 
Assuming that the EW of hydrogen recombination lines are good 
indicators of the age, we find that $20-30\%$ of the detected clusters
in the $H$-band have ages younger than approximately 7 million years. These
young clusters have stellar masses of between $3 \times 10^3\,{\rm 
M}_\odot$ and $3 \times 10^4\,{\rm M}_\odot$.

We finally model the nuclear UV to mid-infrared 
SED of NGC~5253 using the model of Takagi et al. (2003a), and compare
it to the SED of the entire galaxy. For the nuclear SED we have 
taken special care to measure and subtract 
the underlying galaxy emission to isolate 
as much as possible the UV to mid-infrared emission of 
the double star cluster.  For a Salpeter IMF we find that the 
nuclear infrared SED is well reproduced with a total stellar 
mass of $M_{\rm *} = 2.8 \times 10^6\,{\rm M}_\odot$ and an age 
of 3.2 million years in good agreement with the estimates using 
the photometry and spectroscopy of C1+C2. 
The modelling of the nuclear SED shows that the young starburst
is very obscured, $A_V=17\,$mag, in good agreement with 
extinction estimates obtained from near-infrared hydrogen recombination
lines. The UV-optical SED of 
the nuclear region is not well reproduced by the model probably 
due to the complicated dust and emitting source
geometry of the nuclear region. The comparison between 
the nuclear and entire galaxy SED modelling shows that up to 50\%
of the dust mass in this galaxy is concentrated in the nucleus.

The model of the nuclear SED of NGC~5253 
also predicts a moderately bright 
$3.3\,\mu$m PAH feature that is not observed in our 
nuclear $L$-band spectrum. The $3.3\,\mu$m PAH feature along with 
the mid-infrared PAH
features are detected in most massive starburst galaxies, but appear to be 
weak or absent in low mass and metallicity dwarf star forming galaxies. 
Rigby \& Rieke (2004) have recently demonstrated that the star formation
properties of NGC~5253, in particular the mid-infrared fine
structure line
ratios, can only be reproduced with a Salpeter IMF where the upper
mass cutoff is $M_{\rm up} > 40-60\,{\rm M}_\odot$. In the other more massive
and metal rich starbursts in their sample 
the interstellar radiation field may  not include the outputs of stars
above $40\,{\rm M}_\odot$ because most of them are probably embedded in
compact H\,{\sc ii} regions.
The fact that low metallicity stars produce a harder radiation field
together with the presense of massive stars  
in the nuclear starburst of 
NGC~5253 result in a more intense and harder radiation field that 
could effectively suppress the production of the $3.3\,\mu$m PAH feature.

\smallskip

We thank Henrik Spoon for providing us with the $L$-band spectrum of NGC~4945, 
and J. Turner and S. Beck for providing us with the VLA radio
data of NGC~5253. We are also grateful to Mike Cushing for advice
on Spex data reduction, Jason Harris for providing data for 
the optical clusters in electronic format, Karl Gordon for
interesting discussions, and Alice Quillen for assistance 
with the Spex observations. AAH acknowledges support from NASA 
Contract 960785 through the Jet Propulsion Laboratory, and
 the Spanish Programa Nacional de Astronom\'{\i}a y Astrof\'{\i}sica 
under grant AYA2002-01055.

\end{document}